\documentclass{jfm}
\usepackage{graphicx}
\usepackage{epstopdf, epsfig}
\usepackage{subfigure}
\usepackage[dvipsnames]{xcolor}
\usepackage{color}
\usepackage{appendix}
\definecolor{PMcolor}{rgb}{0,0,1}

\shorttitle{Air invasion in compliant microchannels}
\shortauthor{L. Keiser, P. Marmottant and B. Dollet}

\title{Intermittent air invasion in pervaporating compliant microchannels}
\author{Ludovic Keiser\aff{1},
 Philippe Marmottant\aff{1}
 \and Benjamin Dollet\aff{1}
  \corresp{\email{benjamin.dollet@univ-grenoble-alpes.fr}}}

\affiliation{\aff{1}Univ. Grenoble Alpes, CNRS, LIPhy, 38000 Grenoble, France}

\begin{document}

\maketitle

\begin{abstract}

We explore air invasion in an initially water-filled dead-end compliant microchannel containing a constriction. The phenomenon is driven by the pervaporation of the liquid present in the channel through the surrounding medium. The penetration is intermittent, jerky, and characterised by a stop-and-go dynamics as the bubble escapes the constriction. We demonstrate that this sequence of arrest and jump of the bubble is due to an elastocapillary coupling between the air-liquid interface and the elastic medium. When the interface enters the constriction, its curvature strongly increases, leading to a depression within the liquid-filled channel which drives a compression of the channel. As the interface is forced to leave the constriction at a given threshold, due to the ongoing loss of liquid content by pervaporation, the pressure is suddenly released, which gives rise to a rapid propagation of the air bubbles away from the constriction and a restoration of the rest shape of the channel. Combining macroscopic observations and confocal imaging, we present a comprehensive experimental study of this phenomenon. In particular, the effect of the channel geometry on the time of arrest in the constriction and the jump length is investigated.
Our novel microfluidic design succeeds in mimicking the role of inter-vessel pits in plants, which transiently stop the propagation of air embolism during long and severe droughts. It is expected to serve as a building block for further biomimetic studies in more complex leaf-like architectures, in order to recover this universal phenomena of intermittent propagation reported in real leaves.
%\PM{(\citet{Brodribb2016PNAS,Brodribb2016NP} Je ne mettrai pas de citation dans le résumé, il est destiné à être lu séparément)}

\end{abstract}

\begin{keywords}
%Authors should not enter keywords on the manuscript, as these must be chosen by the author during the online submission process and will then be added during the typesetting process (see http://journals.cambridge.org/data/\linebreak[3]relatedlink/jfm-\linebreak[3]keywords.pdf for the full list)
\end{keywords}

\section{Introduction}

Due to climate change, severe droughts are gradually more frequent across the globe. Trees are particularly vulnerable to water stresses, as their sap conduction mainly occurs at negative pressure, close to the cavitation threshold. When air penetrates the vessels of a tree, it forms an embolism which can dramatically impair its hydraulic circulation and lead to its death \citep*{Tyree2013, Venturas2017}, which is one the major threats for the survival of forest for the next decades \citep{Choat2012,Choat2018,Brodribb2020}.

Direct observations of embolism in trees have recently been enabled by the development of high resolution imaging like X-ray tomography \citep*{Cochard2015} or magnetic resonance imaging \citep{Choat2010,Fukuda2015}. However, the time resolution of those methods is generally insufficient to observe fast embolism propagation. Dynamics of embolism growth is more practical to be directly observed in herbaceous plants \citep*{Skelton2017} or in leaves \citep{Brodribb2016NP}, due to their slender geometry and/or their relative transparency. Leaves are particularly exposed to air and more prone to exhibit air embolisation during droughts. A systematic study across various species recently revealed via optical imaging that the drying in real leaves is characterised by an intermittent jerky dynamics \citep*{Brodribb2016PNAS}. These sudden propagations contrast with the smooth flows of sap reported in the absence of air embolisms \citep*{Katifori2010}. Despite a comprehensive description of the embolism formation dynamics in leaves of many species, an in-depth physical understanding is still missing.

In that purpose, we have carried out biomimetic studies, in order to replicate this phenomenon in microfluidic chips. Biomimetic microfluidics in so-called \textit{trees-on-a-chip} has already enabled to reproduce key phenomenon of plant hydrodynamics, like vein circulation at negative pressure \citep{Wheeler2008}, leaf transpiration \citep{Noblin2008}, cavitation in inclusions \citep{Vincent2012, Duan2012, Vincent2017, Bruning2019}, osmotic flows \citep{Jensen2009} or phloem loading \citep{Comtet2017}. Inspired by the design proposed by \cite{Noblin2008} and based on pervaporation across PDMS as a tool to mimic evapotranspiration of leaves, we explored the dynamics of propagation of air embolism in dead-end microchannels \citep{Dollet2019}. However, drying dynamics was found to be smooth, contrary to the phenomenology reported in real leaves. In our previous studies, channel width was taken constant \citep{Dollet2019,Dollet2021}, or slowly varying \citep*{Chagua2021}. In reality, there are pits between sap-conducting cells in trees which are considered to be at the origin of the resistance to embolisation in wood vessels \citep*{Cochard1992}. Those submicron porous membranes separate neighboring vessels and act as passive valves preventing air bubbles to penetrate further in the network due to capillarity. However, under severe drought, the pressure within the hydraulic network can be so low that air embolism manage to pass through the pits and invade very suddenly the network. The compliance of the hydraulic network of trees, i.e. the change of volume of its channels and neighboring cells in response to a pressure change, is considered as an important feature during air embolism formation \citep{Holtta2009}, but live observations are still missing. 

In microfluidics, numerous studies have investigated the deformation of compliant microchannels due to hydrodynamic forces \citep{Holden2003,Gervais2006, Hardy2009, Christov2018} in pure one-phase flows. Furthermore, when two phases are transported in a compliant microchannel, the capillary forces exerted at the fluid-fluid interface can substantially deform the walls, a phenomenon referred to as elasto-capillarity \citep{Roman2010}. It can lead to complex couplings between two-phase flows and compliant walls \citep*{Heil2016,Juel2018}, as for viscous fingering instabilities \citep{Pihler2012,Pihler2013} or for the propagation of air fingers in rectangular compliant microchannels where highly non-linear dynamics were reported \citep{Ducloue2017a,Ducloue2017b,Fontana2021}. Those latter considerations are of primary importance for numerous biological flows \citep{Heil2011}, in particular in collapsed lungs \citep*{Grotberg1994,Grotberg2004,Heil2008}.

In this paper, we consider the physical effects of a single constriction inside a compliant channel of otherwise constant width, in order to mimic the role of pits in plants. The water, initially present in the vessel, gradually pervaporates through the surrounding medium, a phenomenon replicating the effect of evapotranspiration in real leaves. As the air bubble (or air embolism) grows within the channel and penetrates the constriction, we unravel a stop-and-go dynamics, which may be the building block of the intermittent drying dynamics in leaves. We show that this highly non-linear dynamics is generated by the elasto-capillary coupling between the compliance of the channel and the interface curvature, forced to sustain large changes due to the constriction, thereby inducing substantial changes of the capillary pressure within the liquid-filled part of the channel.

In a first part, we describe the experimental materials and methods used in this study (section \ref{part:matetmeth}). We then describe extensively the dynamics of propagation of the air embolism induced by pervaporation, recalling the smooth dynamics in the absence of constriction (section \ref{part:smooth}), contrasting with the stop-and-go induced by the constriction (section \ref{part:stopandgo}). We associate this dynamics with observations of the deformation of the channels via confocal imaging (section \ref{part:confocal}), and show that the deformation results from an elasto-capillary coupling between the meniscus in the constriction and the deformable channel (section \ref{part:elastcapillarity}), in agreement with experiments realised replacing water by ethanol (section \ref{part:ethanol}) and with a volume conservation check comparing macroscopic experiments and confocal imaging (section \ref{part:volume}). We then reinforce the validity of this physical mechanism by depicting the influence of the channel geometry on two key variables of this stop-and-go dynamics: the length of the jump (sections \ref{part:l_out}, \ref{Sec:compliance} and \ref{part:w_p}) and the time of residence of the interface inside the constriction (section \ref{part:t_res}). We also describe the dynamics of the sudden propagation of the air embolism past the constriction, and compare it with theoretical predictions (section \ref{part:dynamics}). In a last part, we summarise our findings and propose connections with other fields as well as perspectives for future studies (section \ref{part:Summary}).

\section{Materials and methods}
\label{part:matetmeth}

\subsection{Photolithography}

The channels are made by photolithography. A SU8 resin (provided by Gersteltec) is spin-coated on a silicon wafer, at a thickness $h$. By photolithography, the negative pattern of the channels is cured by UV-light through a mask (provided by Selba company). The uncured resin is then washed out using propylene glycol methyl ether acetate. After a final bake of the silicon wafer at 200$^\circ$C, our textured surface is ready to use as a mold.

\subsection{PDMS preparation}

In order to make our biomimetic transpirating leaves, we use polydimethylsiloxane (Sylgard$^{TM}$ 184, from Dow company), a silicone-based elastomer. We mix a mass fraction of $10\%$ of curing agent and 90$\%$ of base. After an energic stirring, the mixture is degassed under vacuum. After about $60$ min of degassing, we spin-coat $2.5$ mL of the mixture on the textured silicon surface ($10$ s at $500$ rounds per minute (rpm) and $40$ s at a speed varying between $700$ and $2000$ rpm depending on the desired thickness $H$). The spin-coated liquid film is then let at rest during $1$ h in order to have a homogeneous thickness by capillary levelling. Finally, the samples are put in an oven at $65^\circ$C during $24$ h (or $1$ h depending on the protocol).

\subsection{Channel formation and filling with water}
\label{part:filling}

After curing of the PDMS, the samples are detached from the textured silicon wafer, and gently deposited on a plasma-treated glass slide. The channels are full of air, and we fill them by water using the following procedure. With a scalpel, we create a wide cut enabling a direct connection between the channel entrance and the outside. The samples are then plunged in a tank of deionised water (resistance of about $18$ M$\Omega$) and put in a vacuum chamber during two hours, such that water gradually replaces the air inside the microchannels.

\subsection{Channel design}

\begin{figure}
  \centerline{\includegraphics[width=\columnwidth]{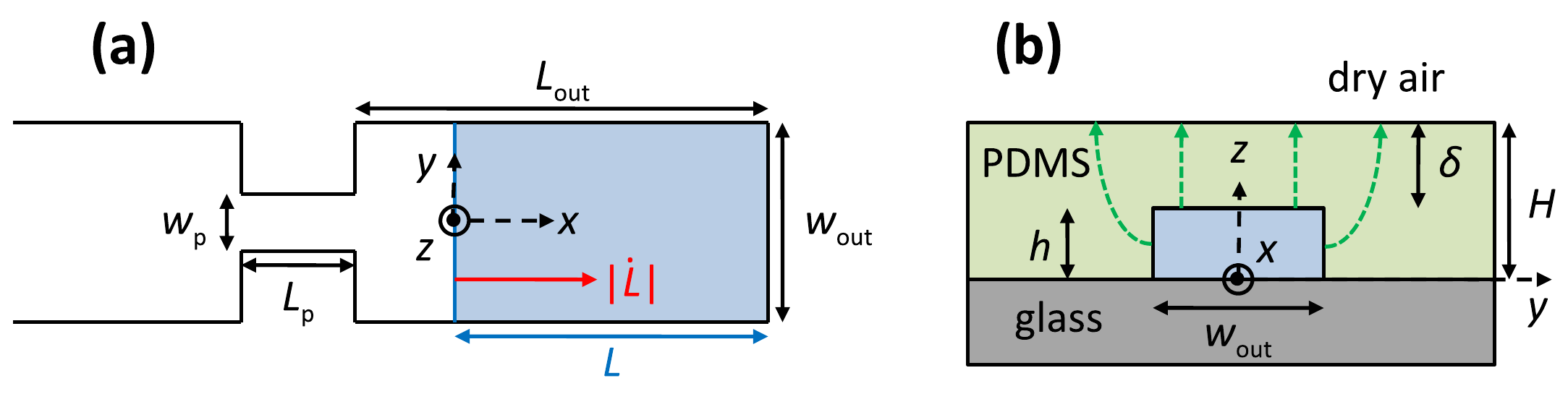}}% Images in 100% size
  \caption{Sketch of the channel (not to scale): (a) top view, (b) side view across the channel. The top view shows the single constriction of length $L_{\mathrm{p}}$ and width $w_{\mathrm{p}}$ inside the channel of width $w_{\mathrm{out}}$. The exit channel, i.e. the channel portion after the constriction, has a length $L_{\mathrm{out}}$. A meniscus, depicted as a blue line across the channel, separates the water-filled part of the channel (right of the meniscus, blue shade in both panels) of length $L$ from the air-filled part (left of the meniscus). The velocity of the meniscus $|\dot{L}|$ is indicated with a red arrow in panel (a); the meniscus motion is due to the loss of water induced by a pervaporation flux across PDMS, sketched by the green dotted arrows in panel (b).}
\label{Fig:Sketch_channel}
\end{figure}

As represented in figure \ref{Fig:Sketch_channel}, our biomimetic vessel is composed of two channels of length $L_{\mathrm{in}}=2$ mm and $L_{\mathrm{out}}$ (varying between $900\:\mathrm{\mu}$m and $18$ cm), of respective width $w_{in}=200\:\mathrm{\mu}$m and $w_{\mathrm{out}}$ (varying between $100$ and $450\:\mathrm{\mu}$m). The two channels have a height $h$ (varying between $11$ and $100\:\mathrm{\mu}$m), and are connected by a constriction of length $l_c = 1$ mm, width $w_{\mathrm{p}}$ (varying between $5$ and $50\:\mathrm{\mu}$m) and height $h$.  The total thickness of the PDMS medium is $H$ and typically varies between $56$ and $154$ $\mu$m across experiments. The thickness of the PDMS membrane separating the top of the channels and the external air is denoted $\delta = H-h$ and varies between $10$ and $79$ $\mu$m across experiments.

\subsection{Meniscus tracking}

In order to record the global displacement of the embolism, we mounted a camera (Pike, from Allied Vision Technologies) on a microscope (magnification varying between $\times 3$ and $\times 5$), and we recorded the position of the meniscus with an acquisition speed varying between 0.1 and 2.5 frames per second (fps).
The position of the meniscus is then tracked using the ``reslice'' function of the free software ImageJ/Fiji. 

In order to record the temporal evolution of the interface at the moment of the sudden jump, we mounted a high speed camera (Phantom Miro M$310$) on a microscope (magnification varying between $\times 6.4$ and $\times 8$) with an acquisition speed varying between $400$ and $1000$ fps.

\subsection{Confocal imaging}

A mass of $1$ mg of sulforhodamine B (a fluorescent dye) was dissolved in 500 mL of deionised water. The solution was then used to fill our channels, following the protocol detailed in section \ref{part:filling}. Fluorescent-water-filled microchannels were put under a laser-induced fluorescence confocal microscope (TCS SP8 from Leica Microsystems), and enlighted at a wavelength $\lambda=561$ nm.
We used a $\times 40$ lens with immersion oil, connected to a photomultiplicator in the wavelength range $[575;\:660]$ nm. The pixel size in our experiments was set at about $200$ nm. We performed $z$-scans of our samples of typical vertical steps $1$ $\mu$m, thereby obtaining images of the channel cross-section (see later figure \ref{Fig:S_t_ref}b) every 22~s. These images (see later figure \ref{Fig:S_t_ref}b) were then analysed by standard image analysis techniques. As a word of caution, notice that sulforhodamine accumulates at the channel boundary and defines a thick bright line, which limits the precision of the measurement of the cross-section and the channel thickness. In particular, we shall see later (in section \ref{part:confocal}) that both these parameters are slightly overestimated, without consequences on their relative variations along time and space, which will be the quantities of interest.

\section{Experimental description of the air-finger propagation} \label{Sec:description_dynamics}

\begin{figure}
    \centering
    \begin{subfigure}[]
        {\includegraphics[width=0.45\textwidth]{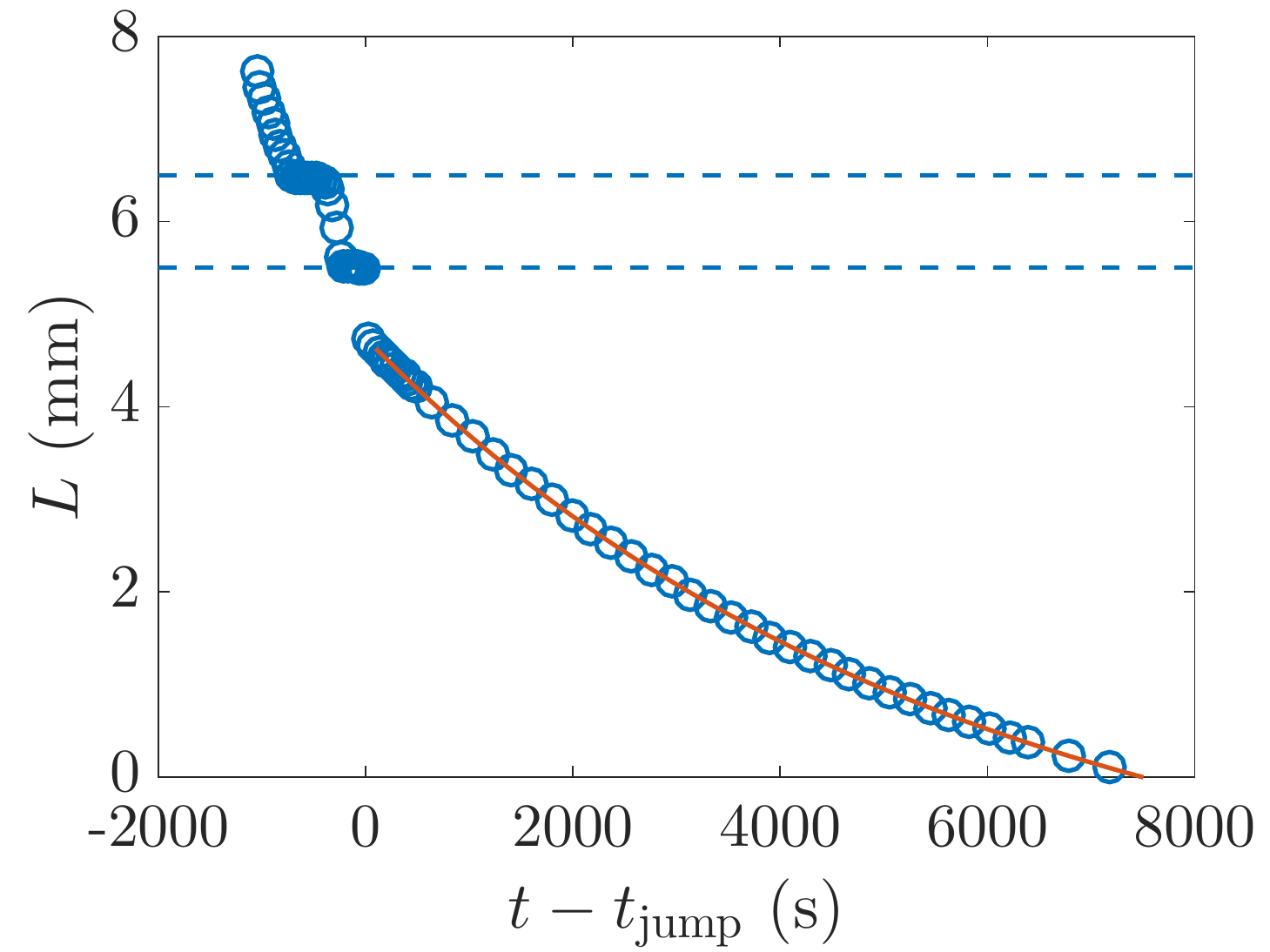}}
        %\caption{$h = 35~\mu$m, $H = 75~\mu$m.}
        %\label{Fig:L(t_tc)_h035um_H075um}
    \end{subfigure}
    ~ %add desired spacing between images, e. g. ~, \quad, \qquad, \hfill etc. 
      ~\begin{subfigure}[]
        {\includegraphics[width=0.45\textwidth]{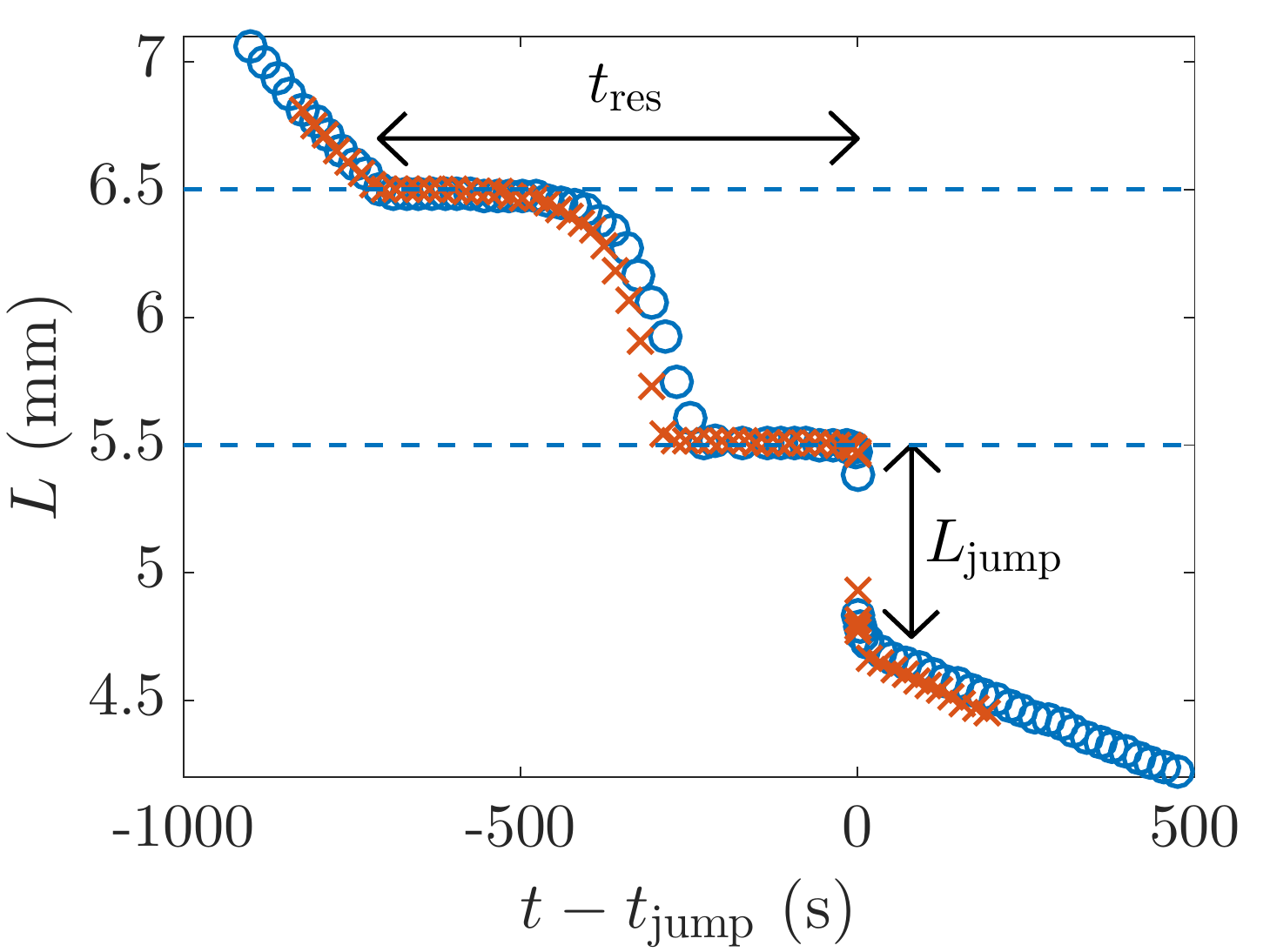}}
        %\caption{$h = 37~\mu$m, $H = 102~\mu$m.}
        %\label{Fig:L(t_tc)_h037um_H102um}
    \end{subfigure}
    
    \begin{subfigure}[]
        {\includegraphics[width=0.6\textwidth]{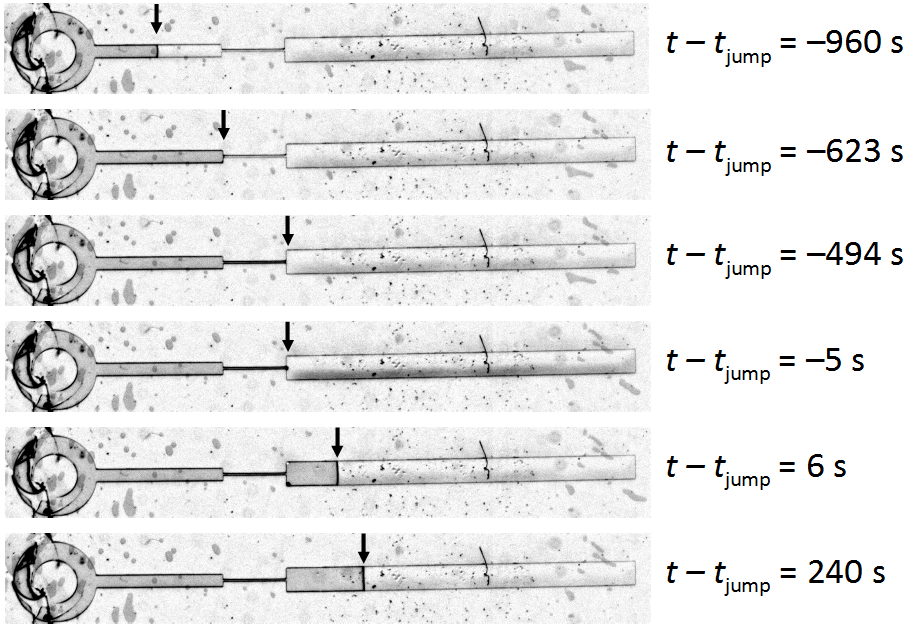}}
        %\caption{$h = 35~\mu$m, $H = 75~\mu$m.}
        %\label{Fig:L(t_tc)_h035um_H075um}
    \end{subfigure}
    \caption{(a) Plot of the water length $L$ as a function of time $t - t_{\mathrm{jump}}$ in a channel of geometrical parameters $L_{\mathrm{out}} = 5.5$~mm, $L_{\mathrm{p}} = 1$~mm, $w_{\mathrm{out}} = 390~\mu$m, $w_{\mathrm{p}} = 30~\mu$m, $h = 65~\mu$m and $H = 95~\mu$m, for a relative humidity $\mathrm{RH} = 0.25$. The time $t_{\mathrm{jump}}$ is the instant where the meniscus abruptly jumps inside the exit channel. The curve is a fit by the truncated exponential law (\ref{Eq:truncated_exponential}), with $\tau = 5.65 \times 10^3$~s and $L_g = 1.71$~mm as fitting parameters. (b) Zoom in the constriction of the experiment presented in panel (a) ($\circ$), and of a second realisation of the experiment ($\times$), recorded with a larger magnification. In panels (a) and (b), the two horizontal dashed lines highlight the location of the constriction. (c) Snapshots of the full channel at various times. Arrows highlight the position of the meniscus separating the water-filled part of the channel (right from the meniscus) from its air-filled part (left from the meniscus). %\textcolor{red}{1. EITHER: put a supplemental figure showing a wider view of the constriction plus meniscus, with a rectangle showing the zoom we make to perceive the four (i) (ii) (iii) and (iv) vizualisations of the mensicus. It might be a bit hard for now to understand clearly what we are seeing. 2. OR: Put an intermediary figure (future new figure 2) showing a full trajectory L(t) with L reaching 0, and show at different key moments the global view of the channel.}
    } \label{Fig:L_t_ref}
\end{figure}

We start by considering a reference experiment, whose dynamics is representative of that of all other experiments presented in this paper. The time evolution of the water length $L$ is plotted in figure \ref{Fig:L_t_ref}. It shows sharp variations: the meniscus first advances at almost constant speed in the entrance channel, before reaching the constriction entrance, where it stops for about $200$ s. It then accelerates inside the constriction, before a second stop at the constriction exit, for about $250$ s. It then suddenly jumps forwards over a length $L_{\mathrm{jump}}$ of about 1~mm inside the exit channel, at a time denoted $t_{\mathrm{jump}}$, before recovering a smooth dynamics in the exit channel characterised by a progressive deceleration. We supply as Supplementary Material a movie showing the whole process. We study more in details this smooth dynamics in section \ref{part:smooth}, then focus on the constriction-induced sharp variations in section \ref{part:stopandgo} and propose physical interpretations in sections \ref{part:confocal}, \ref{part:elastcapillarity}, \ref{part:ethanol} and \ref{part:volume}.

%\begin{figure}
%  \centerline{\includegraphics[width=0.5\columnwidth]{Dynamique_globale.pdf}}% Images in 100% size
%  \caption{[RAJOUTER DEUX TRAITS HORIZONTAUX POUR L'ENTR\'EE ET LA SORTIE DE LA CONSTRICTION] Plot of the water length $L$ as a function of time $t - t_{\mathrm{jump}}$ in a channel of geometrical parameters $L_{\mathrm{out}} = 5.5$~mm, $L_p = 1$~mm, $w_{\mathrm{out}} = 390~\mu$m, $w_p = 30~\mu$m, $h = 65~\mu$m and $H = 95~\mu$m. The time $t_{\mathrm{jump}}$ is the instant where the meniscus abruptly jumps inside the exit channel. The curve is a fit by the truncated exponential law (\ref{Eq:truncated_exponential}), with $\tau = 5.65 \times 10^3$~s and $L_g = 1.71$~mm as fitting parameters [COMPARER CES VALEURS \`A CELLES CIT\'EES DANS NOS TROIS PR\'EC\'EDENTS ARTICLES~?].}
%\label{Fig:Dynamique_globale}
%\end{figure}

\subsection{A smooth dynamics in the absence of constriction}
\label{part:smooth}

The smooth deceleration of the meniscus occurring after the jump in the exit channel echoes the pervaporation-induced drying dynamics in single channels of constant width studied by \citet{Dollet2019}. We remind that this dynamics is driven by the loss of water within the channel due to pervaporation through the water-permeable PDMS walls. There are two contributions to the pervaporation flux $Q$: a diffusion flux, denoted $Q_\ell$, from the water--PDMS interface, and an evaporative flux $Q_g$ at the meniscus. Since the length of the exit channel is much larger than its width, it is reasonable to neglect end effects on the diffusion flux and to assume that it is simply proportional to the water-filled length: $Q_\ell = q_\ell L$. Hence, water conservation leads to: $hw_{\mathrm{out}} \dot{L} = -q_\ell L - Q_g$, from which the following truncated exponential dynamics is predicted:
\begin{equation} \label{Eq:truncated_exponential}
    L = (L_{\mathrm{out}} - L_{\mathrm{jump}} + L_g) \mathrm{e}^{-(t - t_{\mathrm{jump}})/\tau} - L_g ,
\end{equation}
using the initial condition $L = L_{\mathrm{out}} - L_{\mathrm{jump}}$ at $t = t_{\mathrm{jump}}$ just after the jump, and defining $\tau = hw_{\mathrm{out}}/q_\ell$ and $L_g = Q_g/q_\ell$. Figure \ref{Fig:L_t_ref}a confirms that such a truncated exponential fits remarkably well the meniscus dynamics in the exit channel. A similar dynamics is very probably at play also at the entrance channel, but the spatiotemporal range of this part of the dynamics is too narrow to enable for significant fitting.

The two aforementioned fluxes $q_\ell$ and $Q_g$ have been predicted in \citet{Dollet2019}. First, we have:
\begin{equation}\label{Eq:ql}
	q_\ell = D_P \bar{c}_P^{\mathrm{sat}} (1 - \mathrm{RH}) \tilde{q}_\ell(w_{\mathrm{out}}) ,
\end{equation}
with $D_P$ the diffusivity of water in PDMS, $\bar{c}_P^{\mathrm{sat}} = 7.2 \times 10^{-4}$ the mass fraction of water in PDMS at saturation (see \citet*{Chagua2021} for the numerical estimation of this parameter), RH the relative humidity of the outer air, and $\tilde{q}_\ell$ the following dimensionless shape factor:
\begin{equation}\label{Eq:ql_tilde}
	\tilde{q}_\ell(w) = \frac{w}{\delta} + \frac{2}{\pi} \left[ \ln\frac{(H + \delta)h}{\delta^2} + \frac{H}{\delta} \ln\frac{H + \delta}{h} \right] .
\end{equation}
We also use the prediction:
\begin{equation}\label{Eq:Qg}
	Q_g = \sqrt{\alpha D_a D_P hw_{\mathrm{p}} \tilde{q}_\ell(w_{\mathrm{out}})} ,
\end{equation}
with $\alpha = 0.03$ the Henry constant quantifying the water affinity in PDMS \citep*{Harley2012}, and $D_a = 2\times 10^{-5}$~m$^2$/s the diffusivity of water vapour in air.

\subsection{A stop-and-go dynamics induced by the constriction}
\label{part:stopandgo}

Figure \ref{Fig:L_t_ref}b shows that the presence of the constriction modifies dramatically the drying dynamics as compared to the smooth dynamics encountered in channels of constant width \citep{Dollet2019}. Moreover, the dynamics is also different to the one studied in \citet*{Chagua2021} for channels of slowly varying width, where the width variations were shown to induce variations of meniscus speed, but not the stick-slip-like motion (arrest followed by jump) evidenced in figure \ref{Fig:L_t_ref}b.

%We start by considering a reference case, which is representative of all our experiments. The time evolution of the water length is plotted in figure \ref{Fig:L_t_ref}. It shows sharp variations: the meniscus first advances at almost constant speed before reaching the constriction entrance, where it accelerates, before getting arrested at the constriction exit. After a certain arrest time (of about 500~s), it suddenly jumps forwards inside the exit channel (over a length of about 1~mm) before recovering a smooth dynamics. Hence, the presence of the constriction modifies dramatically the drying dynamics as compared to the smooth dynamics encountered in channels of constant width \citep{Dollet2019}. Moreover, the dynamics is also different to the one studied in \citet{Chagua2021} for channels of slowly varying width, where the width variations were shown to induce variations of meniscus speed, but not the stick-slip-like motion (arrest followed by jump) evidenced in figure \ref{Fig:L_t_ref}.

%\begin{figure}
%  \centerline{\includegraphics[width=0.5\columnwidth]{L_t_ref_2.pdf}}% Images in 100% size
%  \caption{Plot of the water length $L$ as a function of time $t - t_{\mathrm{jump}}$ in a channel of geometrical parameters $L_{\mathrm{out}} = 5.5$~mm, $L_p = 1$~mm, $w_{\mathrm{out}} = 390~\mu$m, $w_p = 30~\mu$m, $h = 50~\mu$m and $H = 59~\mu$m. The time $t_{\mathrm{jump}}$ is the instant where the meniscus abruptly jumps inside the exit channel. The two symbols $\circ$ and $\times$ stand for two independent realisations of the experiment.}
%\label{Fig:L_t_ref}
%\end{figure}

\begin{figure}
    \centering
    \begin{subfigure}[]
        {\includegraphics[width=0.28\textwidth]{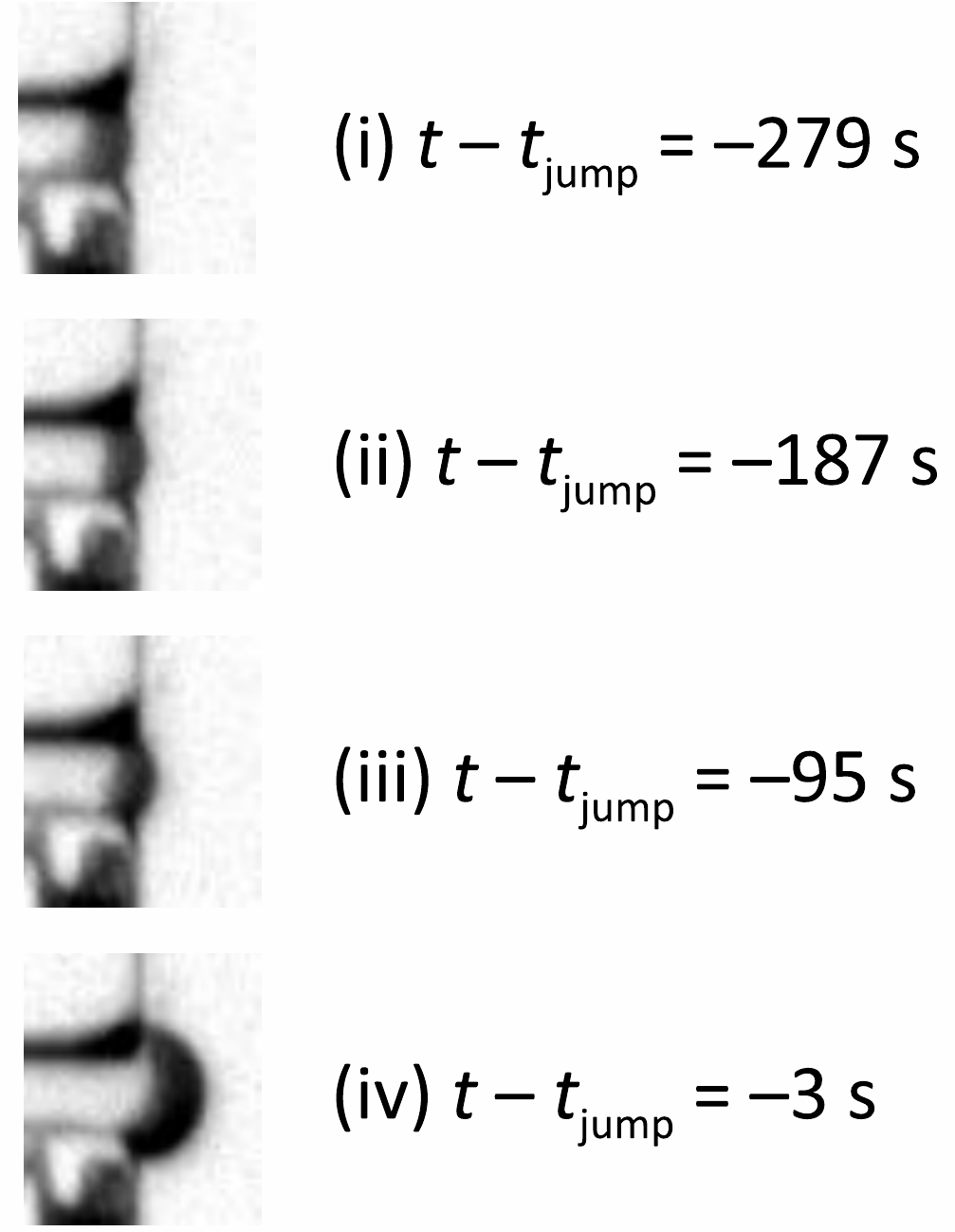}}
        %\caption{$h = 35~\mu$m, $H = 75~\mu$m.}
        %\label{Fig:L(t_tc)_h035um_H075um}
    \end{subfigure}
    ~ %add desired spacing between images, e. g. ~, \quad, \qquad, \hfill etc. 
      ~\begin{subfigure}[]
        {\includegraphics[width=0.5\textwidth]{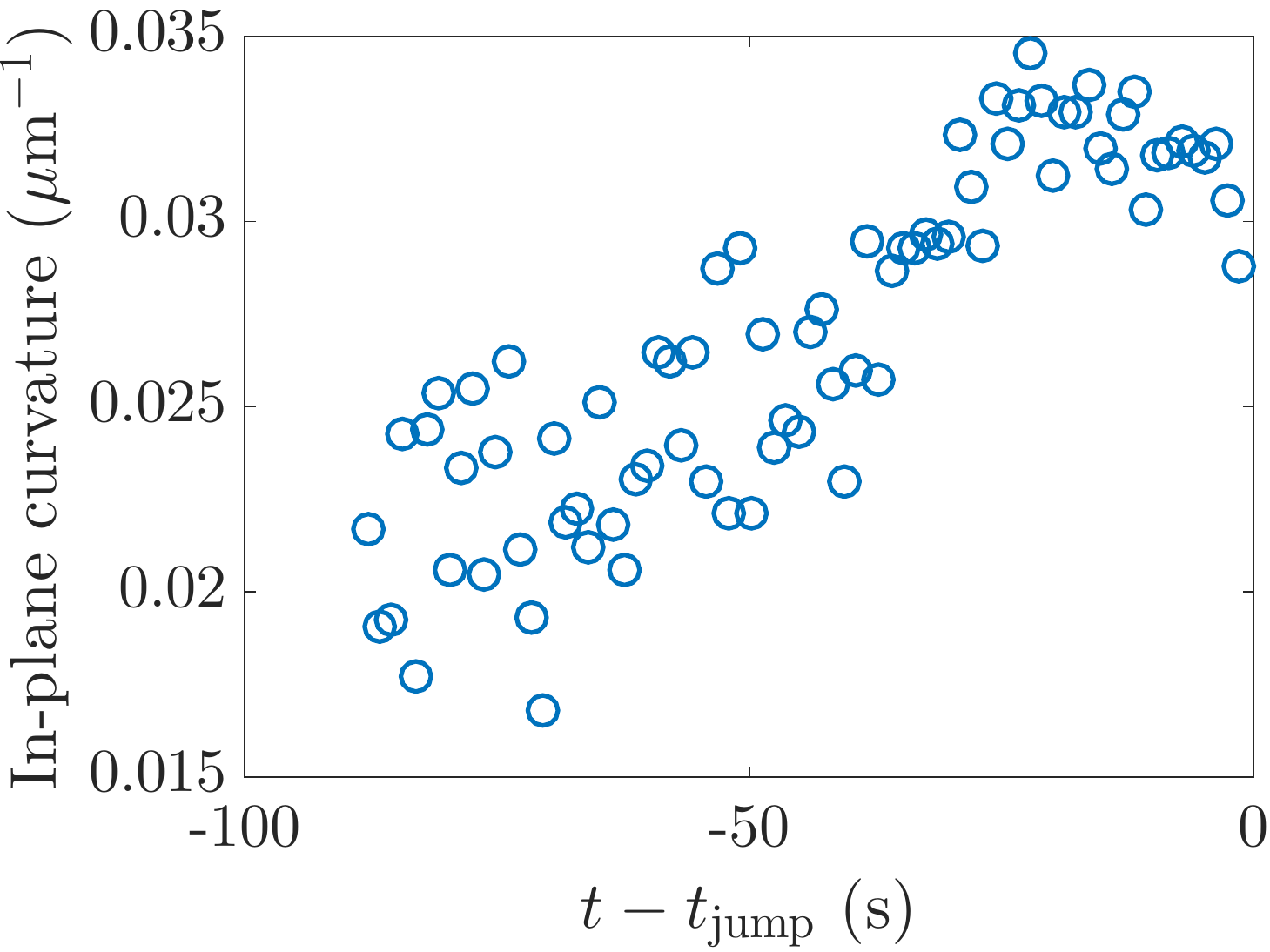}}
        %\caption{$h = 37~\mu$m, $H = 102~\mu$m.}
        %\label{Fig:L(t_tc)_h037um_H102um}
    \end{subfigure}
    \caption{(a) Snapshots of the meniscus at various times within the phase of arrest of the meniscus at the constriction exit; in particular, (i) is the first snapshot where the meniscus is pinned. (b) Plot of the in-plane curvature of the meniscus as a function of time at the end of the arrest phase. %\textcolor{red}{Peut-être qu'on pourrait mettre l'axe des $y$  à 0 afin de ne pas exagérer la perception qu'on peut avoir du bruit.} %\textcolor{red}{1. EITHER: put a supplemental figure showing a wider view of the constriction plus meniscus, with a rectangle showing the zoom we make to perceive the four (i) (ii) (iii) and (iv) vizualisations of the mensicus. It might be a bit hard for now to understand clearly what we are seeing. 2. OR: Put an intermediary figure (future new figure 2) showing a full trajectory L(t) with L reaching 0, and show at different key moments the global view of the channel.}
    } \label{Fig:meniscus_curvature}
\end{figure}

We must mention that some elements of this dynamics are not fully reproducible when the same experiment is run several times. Notably, the arrest times at the constriction entrance and exit, as well as the dynamics within the constriction, vary, as evidenced by the comparison of two different realisations of the experiment in figure \ref{Fig:L_t_ref}b. However, the occurrence of the jump at the constriction exit was always observed, and the jump length and arrest time both showed a good reproducibility (of the order of 10\%).

To understand the origin of the meniscus arrest, we present magnified snapshots of the meniscus in figure \ref{Fig:meniscus_curvature}a at the constriction exit, which show that the meniscus progressively bulges out during the arrest phase. To quantify this observation, we fitted the meniscus by an arc of circle, and we plot in figure \ref{Fig:meniscus_curvature}b the corresponding in-plane curvature as a function of time during the last 90~s before the jump (at earlier times, the meniscus is not curved enough for its curvature to be accurately measurable): although the measurement is somewhat noisy, this figure shows that the curvature increases and reaches a maximum just before the jump, of order 0.035~$\mu$m$^{-1}$, which is very close to $1/w_{\mathrm{p}} = 0.033~\mu$m$^{-1}$.

%\begin{figure}
%  \centerline{\includegraphics[width=0.5\columnwidth]{Courbure_vs_temps_manip_zoomee.pdf}}% %Images in 100% size
%  \caption{Plot of the in-plane curvature of the meniscus as a function of time at the end of the arrest phase. \textcolor{red}{Peut-être qu'on pourrait mettre l'axe des $y$  à 0 afin de ne pas exagérer la perception qu'on peut avoir du bruit.}}
%\label{Fig:Courbure_temps}
%\end{figure}

While the meniscus gets arrested, drying is still proceeding, hence the volume of water in the channel is still decreasing. This suggests that the channel cross-section, in its water-filled part, must decrease during the arrest phase.

\subsection{Confocal imaging of the deformation of the channel during the arrest phase}
\label{part:confocal}

\begin{figure}
    \centering
    \begin{subfigure}[]
        {\includegraphics[width=0.4\textwidth]{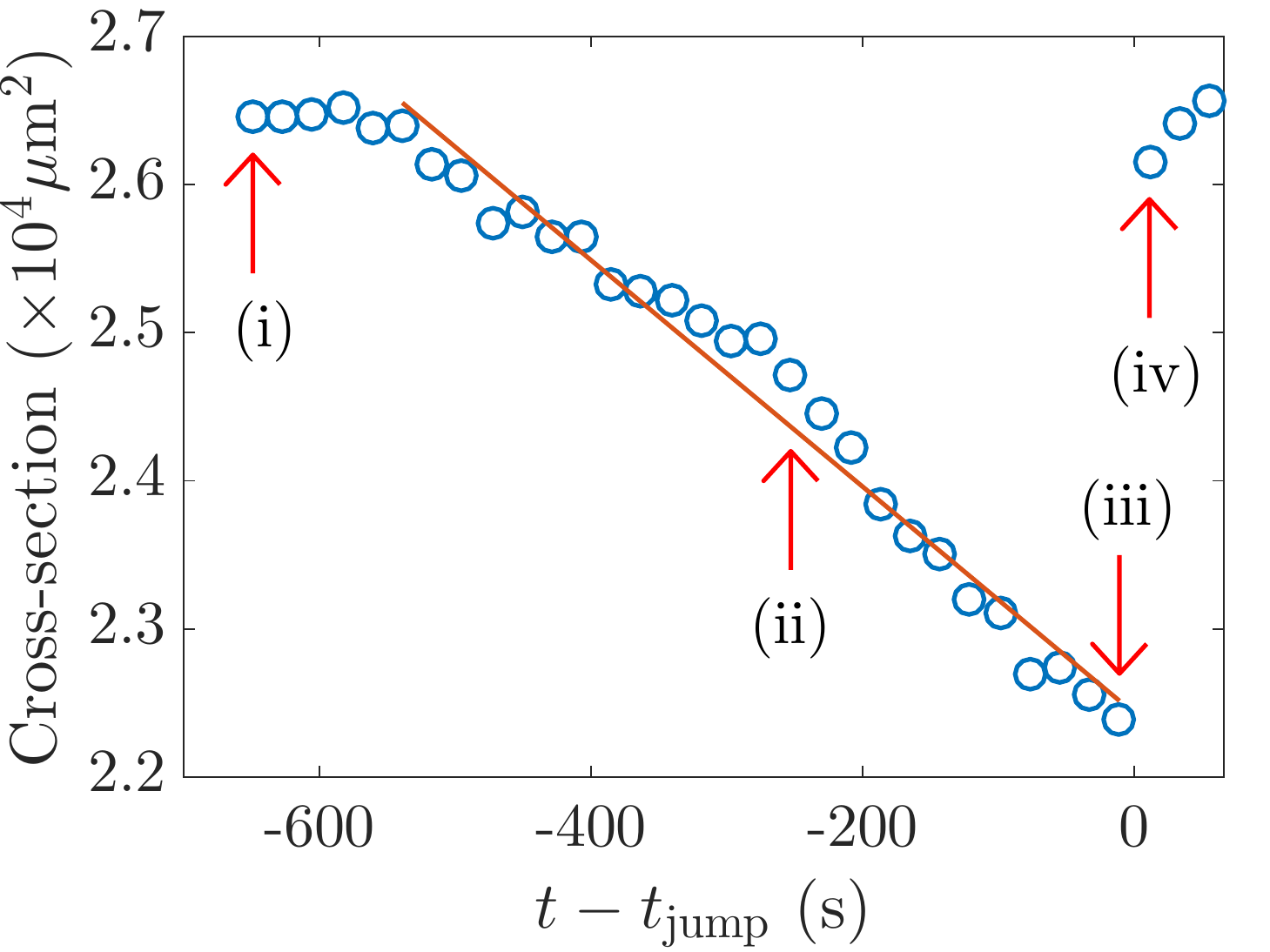}}
        %\caption{$h = 35~\mu$m, $H = 75~\mu$m.}
        %\label{Fig:L(t_tc)_h035um_H075um}
    \end{subfigure}
    ~ %add desired spacing between images, e. g. ~, \quad, \qquad, \hfill etc. 
      ~\begin{subfigure}[]
        {\includegraphics[width=0.5\textwidth]{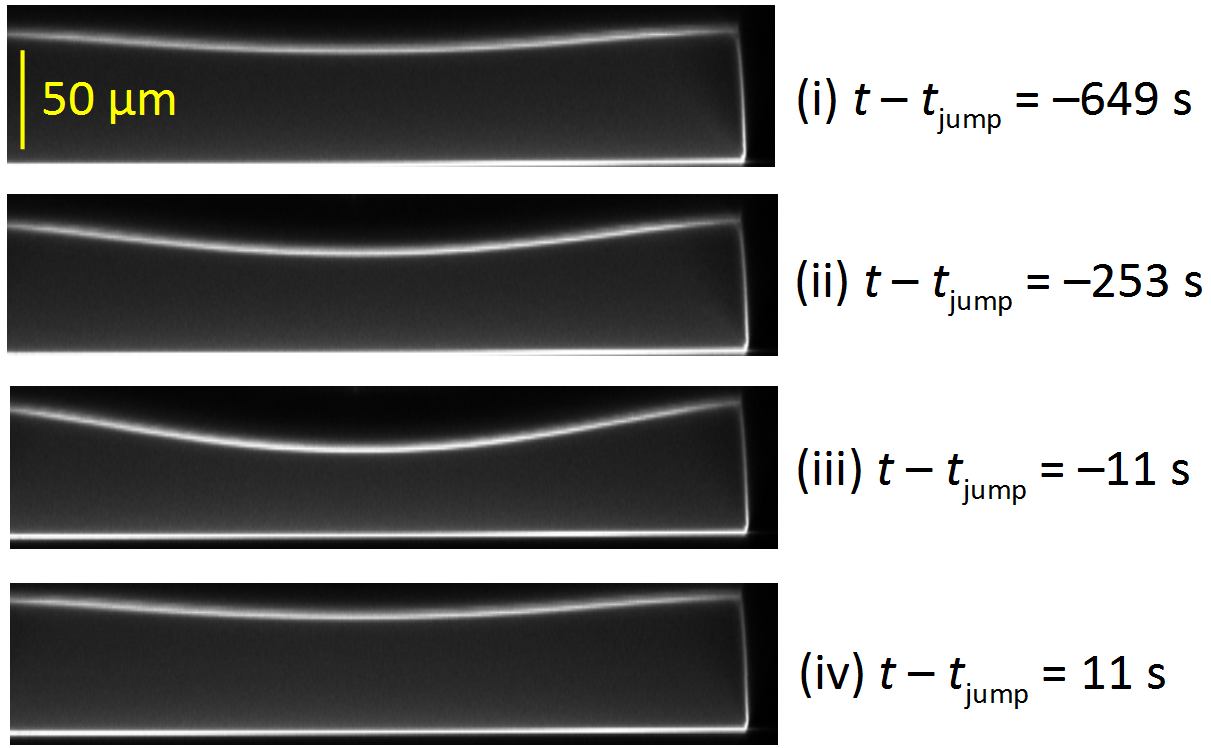}}
        %\caption{$h = 37~\mu$m, $H = 102~\mu$m.}
        %\label{Fig:L(t_tc)_h037um_H102um}
    \end{subfigure}
    \caption{%[PETIT SOUCI~: $S_0 \equiv hw_{\mathrm{out}} = 65 \times 390 = 2.535 \times 10^4~\mu$m$^2$ EST PLUS PETIT QUE LES VALEURS PLATEAU MESUR\'EES, ALORS QU'EN RAISON DE LA PRESSION R\'EMANANTE, IL DEVRAIT \^ETRE PLUS GRAND. POURQUOI~?]\textcolor{red}{[ADDITIVE BIAS DUE TO THE WICKING OF THE FLUORESCENT SIGNAL, IN PARTICULAR IN Z. BY THE WAY, THE HEIGHT OF THE CHANNEL SHOULD BE PUT AT ITS REAL VALUE IN FIGURE 6]} 
    (a) Plot of the cross-section of the channel $S$ as a function of time in a channel of geometrical parameters $L_{\mathrm{out}} = 5.5$~mm, $L_{\mathrm{p}} = 1$~mm, $w_{\mathrm{out}} = 390~\mu$m, $w_{\mathrm{p}} = 30~\mu$m, $h = 65~\mu$m and $H = 95~\mu$m. The time here is defined with respect to the instant of the jump $t_{\mathrm{jump}}$. The straight line is a linear fit over the time interval where the cross-section decreases. Arrows with labels refer to the photographs taken with the confocal microscope displayed in panel (b). The boundary of the channel cross-section appears in light shade; notice that a tiny part of the cross-section is cut at the left of the photographs, owing to limitations of the horizontal field of view of the confocal microscope.} \label{Fig:S_t_ref}
\end{figure}

In order to test the hypothesis of a decrease of the cross-section during the arrest phase, we repeated the reference experiment using confocal microscopy to measure the time evolution of the cross-section. To avoid the end effects in the vicinity of the sharp width variation at the constriction exit, we measured the cross-section at a distance 3~mm ahead of the constriction exit. The corresponding data, displayed in figure \ref{Fig:S_t_ref}a, show that the cross-section is first roughly constant, then decreases, then increases very rapidly (within the 22~s separating two measurements of the cross-section) to recover the first plateau value. This variation is compatible with the aforementioned dynamics of the meniscus, and with a scenario of drying of the channel at fixed meniscus position but decreasing cross-section, until a critical instant where the meniscus unpins. A fast relaxation ensues, during which the channel quickly recovers its equilibrium cross-section, and where the meniscus jumps forwards at almost constant water volume (neglecting the duration of the jump with respect to the typical pervaporation time).

The confocal snapshot of figure \ref{Fig:S_t_ref}b,iv shows that the upper wall remains slightly bent inwards after the jump. This is a signature of a remanent Laplace pressure $p_{\mathrm{rem}}$ when the meniscus is in the exit channel: owing to the wetting conditions of water on PDMS and glass, the meniscus is curved towards the liquid phase (see later figure \ref{Fig:ethanol}b). Hence, the water has a lower pressure than air, whence the remaining curvature of the upper wall. We shall henceforth denote $S_{\mathrm{rem}}$ the corresponding cross-section, which is thus slightly smaller (of order 5\%) than the equilibrium cross-section $S_0 = hw_{\mathrm{out}}$.

%\begin{figure}
%  \centerline{\includegraphics[width=0.5\columnwidth]{S_t_ref.pdf}}% Images in 100% size
%  \caption{Plot of the cross-section of the channel $S$ as a function of time in a channel of geometrical parameters $L_{\mathrm{out}} = 5.5$~mm, $L_p = 1$~mm, $w_{\mathrm{out}} = 390~\mu$m, $w_p = 30~\mu$m, $h = 65~\mu$m and $H = 95~\mu$m. The time here is defined with respect with the instant of the jump $t_{\mathrm{jump}}$.}
%\label{Fig:S_t_ref}
%\end{figure}

\begin{figure}
  \centerline{\includegraphics[width=0.5\columnwidth]{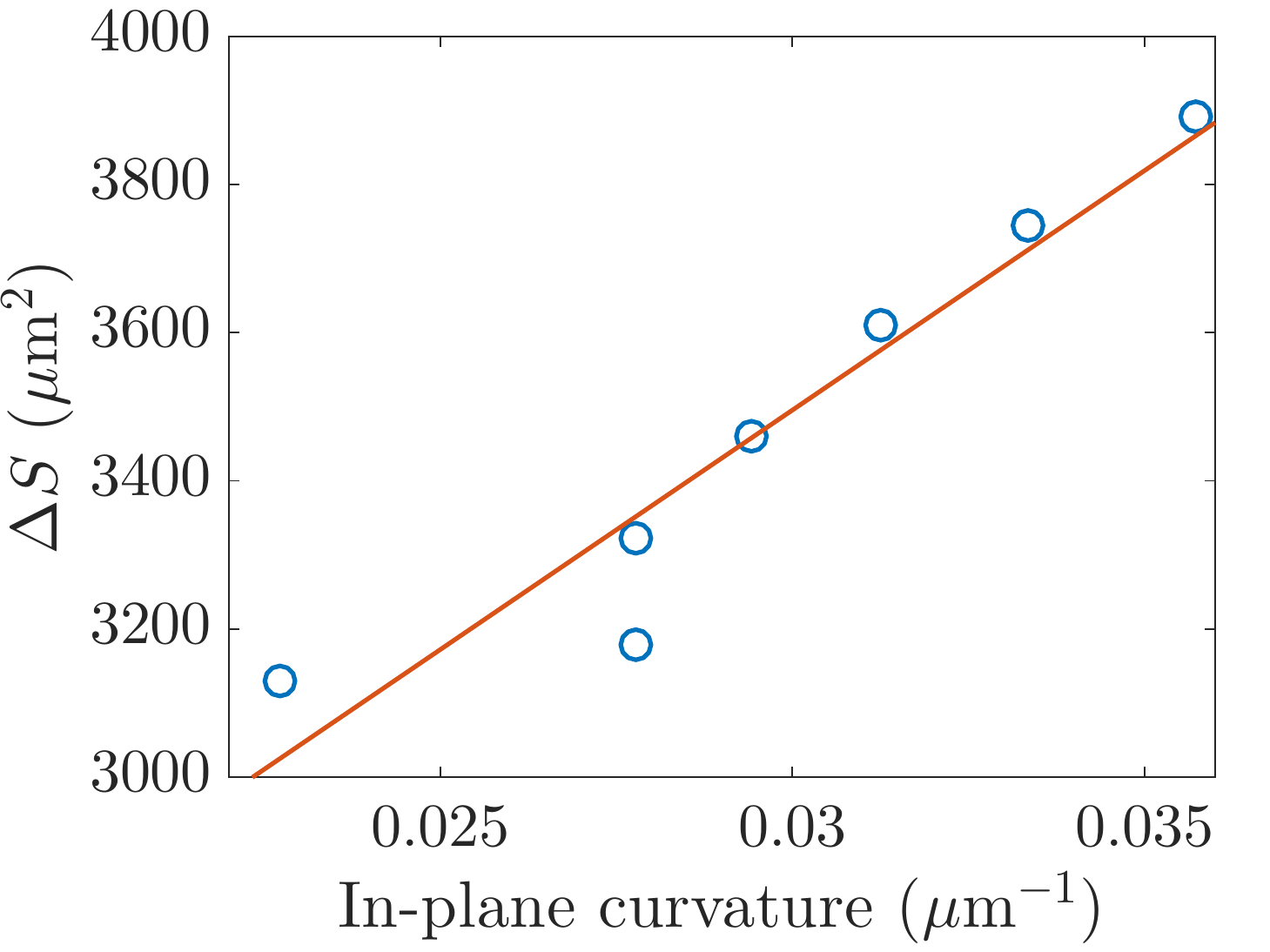}}% Images in 100% size
  \caption{Plot of the cross-section difference of the channel $\Delta S = hw - S$ as a function of the in-plane curvature $\kappa$ of the meniscus in a channel of geometrical parameters $L_{\mathrm{out}} = 5.5$~mm, $L_{\mathrm{p}} = 1$~mm, $w_{\mathrm{out}} = 390~\mu$m, $w_{\mathrm{p}} = 30~\mu$m, $h = 50~\mu$m and $H = 59~\mu$m, for the last moments of the arrest phase. The line is an affine fit of the data $\Delta S = c_1 \kappa + c_2$, with fitting parameters $c_1 = 6.5 \times 10^4~\mu$m$^3$ and $c_2 = 1.6 \times 10^3~\mu$m$^2$.}
\label{Fig:DeltaS_courbure_ref}
\end{figure}

The data of figures \ref{Fig:meniscus_curvature}b and \ref{Fig:S_t_ref}a suggest that the curvature of the pinned meniscus varies in order for the Laplace pressure to accommodate the decrease in water pressure accompanying the cross-section reduction. To prove this correlation, we performed another experiment based on confocal microscopy, close enough to the constriction exit to simultaneously track both the meniscus shape and the cross-section. Owing to the finite time resolution of the confocal scans, we cannot reconstruct reliably the 3D structure of the meniscus, because it advances significantly between the bottom scan and the top scan. Hence, we present a measurement of the curvature of the meniscus in the horizontal plane located midway between the bottom glass plate and the top PDMS channel wall (henceforth called the in-plane curvature). We plot in figure \ref{Fig:DeltaS_courbure_ref} the cross-section difference with respect to the undeformed case, defined as $\Delta S = hw_{\mathrm{out}} - S$, as a function of the in-plane curvature in the last moments of the arrest phase. This figure shows that, except for an outlier, the two quantities are well correlated by an affine relation, thereby confirming our interpretation.

Once the meniscus reaches its maximal possible curvature compatible with the geometrical constraints and the contact angles of water on PDMS and glass, the situation becomes unstable, because the Laplace pressure can no longer equilibrate the elastic deformation of the channel, and the meniscus advances rapidly until the channel is fully relaxed. The onset of the jump is therefore an occurrence of a capillary threshold, as encountered in the drainage of Newtonian fluids in porous media \citep{Wilkinson1983,Lenormand1985,Maloy1992,Meheust2002}, or in the flow of foams in porous media \citep{Geraud2016}. Hence, the maximal deformation of the channel is correlated, through Laplace pressure, to the maximal curvature of the meniscus. Because of the finite roundness of the corners at the constriction exit, the meniscus shape is a complicated surface, which prevents us from predicting the maximal curvature. However, its order of magnitude is $1/w_{\mathrm{p}}$ (figure \ref{Fig:meniscus_curvature}b), as would be expected for a 2D meniscus pinned on sharp corners, whose maximal possible curvature equals $2/w_{\mathrm{p}}$ when it has the shape of a half circle.

\subsection{An elasto-capillary interaction between the meniscus and the channel structure}
\label{part:elastcapillarity}

\begin{figure}
    \centering
    \begin{subfigure}[]
        {\includegraphics[width=0.45\textwidth]{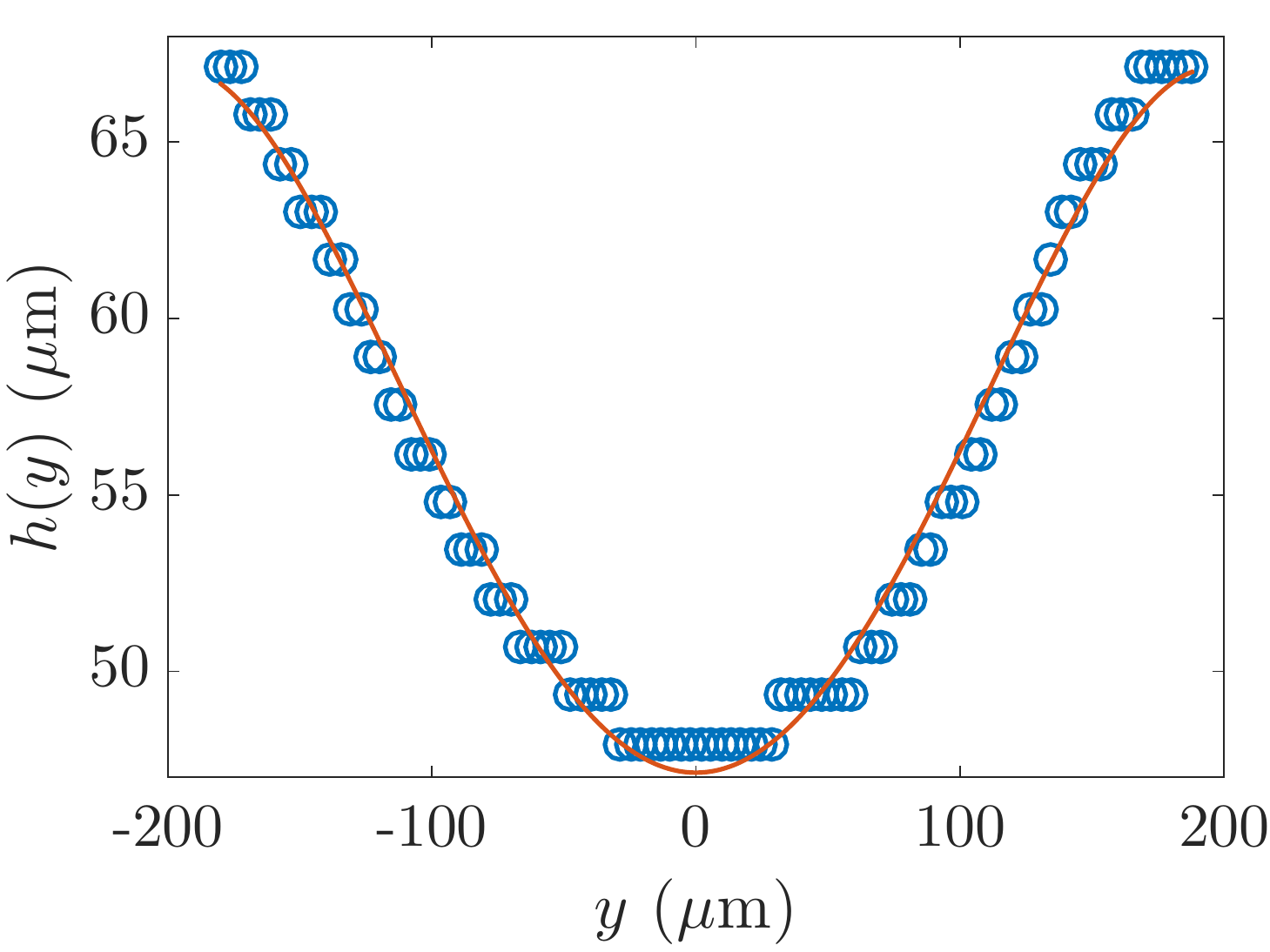}}
        %\caption{$h = 35~\mu$m, $H = 75~\mu$m.}
        %\label{Fig:L(t_tc)_h035um_H075um}
    \end{subfigure}
    ~ %add desired spacing between images, e. g. ~, \quad, \qquad, \hfill etc. 
      ~\begin{subfigure}[]
        {\includegraphics[width=0.45\textwidth]{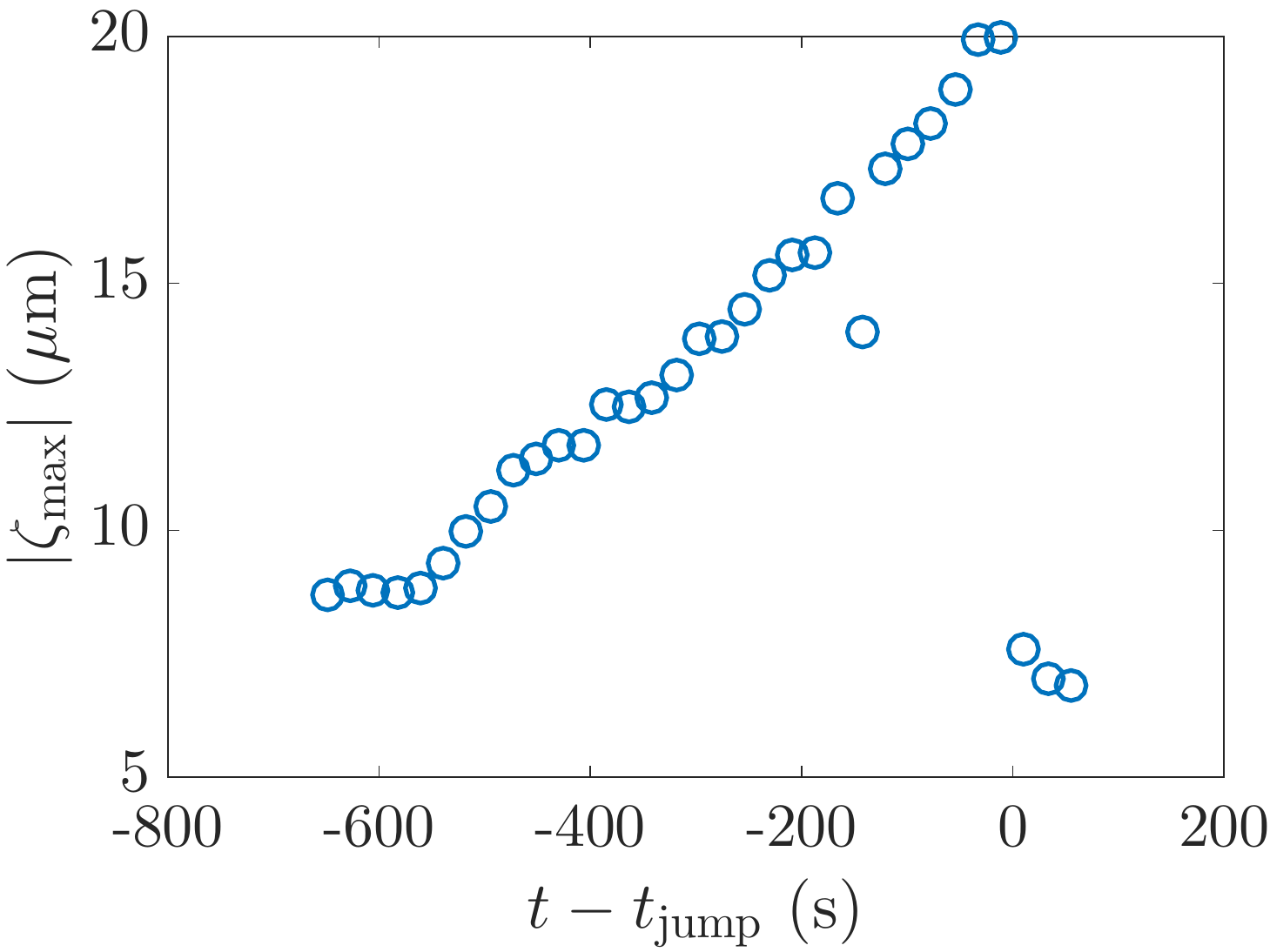}}
        %\caption{$h = 37~\mu$m, $H = 102~\mu$m.}
        %\label{Fig:L(t_tc)_h037um_H102um}
    \end{subfigure}
    \caption{(a) Plot of the local thickness $h(y)$ of the channel from the photograph (iii) in figure \ref{Fig:S_t_ref}b as a function of the spanwise coordinate $y$; the staircase-like rendering of the data points is due to the limited spatial resolution of the confocal microscope. The curve is a fit by the formula $h(y) = h - |\zeta_{\max}|(1 - 4y^2/w_{\mathrm{out}}^2)$ with $|\zeta_{\max}|$ a free fitting parameter, here equal to 20.0~$\mu$m. (b) Plot of $|\zeta_{\max}|$ as a function of time.} \label{Fig:deflection}
\end{figure}

We shall now analyse the full shape of the channel upper wall as seen in figure \ref{Fig:S_t_ref}b. In that experiment, the upper wall has a thickness $\delta = 30~\mu$m and a width $w_{\mathrm{out}} = 390~\mu$m. Hence, it is quite slender and we may compare its deflection to that of a thin beam. The conditions at the boundaries of the upper wall are complex because it is attached to a thicker part of the same material; for the sake of simplicity, we simply assimilate them to clamped end conditions, as assumed in other works \citep{Christov2018,Martinez-Calvo2020}. The deflection of the upper wall is then predicted to equal \citep{Landau1986}:
\begin{equation} \label{Eq:deflection}
    \zeta = \frac{\Delta P}{24B} \left( y^2 - \frac{1}{4} w_{\mathrm{out}}^2 \right)^2 ,
\end{equation}
with $\Delta P$ the pressure difference between the water in the channel and the atmosphere; here, $\Delta P < 0$, hence $\zeta < 0$, corresponding to an inwards deflection of the upper wall. The parameter $B = E\delta^3/[12(1 - \nu^2)]$ is the bending stiffness of the upper wall, with $\nu = 0.498$ the Poisson's ratio of PDMS. To test this prediction, we fit the deflection measured on the last confocal photograph before the jump (figure \ref{Fig:S_t_ref}b,iii). Figure \ref{Fig:deflection}a shows a good agreement between the data point and the predicted deflection. We checked that the agreement remains good on the other confocal photographs, and we plot the maximal deflection $|\zeta_{\max}| = |\Delta P|B/(384B)$, reached at the middle of the upper wall, as a function of time in figure \ref{Fig:deflection}b. It increases linearly with time, before a sudden relaxation after the jump. Moreover, from the maximal deflection obtained just before the jump, we get the numerical value: $(1 - \nu^2)\Delta P/E = 1.0 \times 10^{-3}$. Taking $E = 2$~MPa the Young's modulus of the PDMS, this estimate means that $|\Delta P|$ reaches 2~kPa just before the jump. This estimate agrees with the order of magnitude of the capillary threshold: from figure \ref{Fig:meniscus_curvature}b, the maximal in-plane curvature of the meniscus is about $1/w_{\mathrm{p}}$, whence a maximal Laplace pressure of order $\gamma/w_{\mathrm{p}} = 2$~kPa, where $\gamma = 7\times 10^{-2}$~kg/s$^2$ is the surface tension of water.

%\begin{figure}
%  \centerline{\includegraphics[width=0.5\columnwidth]{Deflexion_toit.pdf}}% Images in 100% size
%  \caption{Plot of the local thickness $h(y)$ of the channel from the photograph (iii) in figure \ref{Fig:S_t_ref}b as a function of the spanwise coordinate $y$; the staircase-like rendering of the data points is due to the limited spatial resolution of the confocal microscope. The curve is a fit by the formula $h(y) = h - c(y^2 - \frac{1}{4} w_{\mathrm{out}}^2)$ with $c$ a free fitting parameter, here equal to $c = 20.0~\mu$m [TRACER $c$ EN FONCTION DE $t$~?].}
%\label{Fig:deflection}
%\end{figure}

\subsection{Influence of surface tension}
\label{part:ethanol}

\begin{figure}
  \centerline{\includegraphics[width=1\columnwidth]{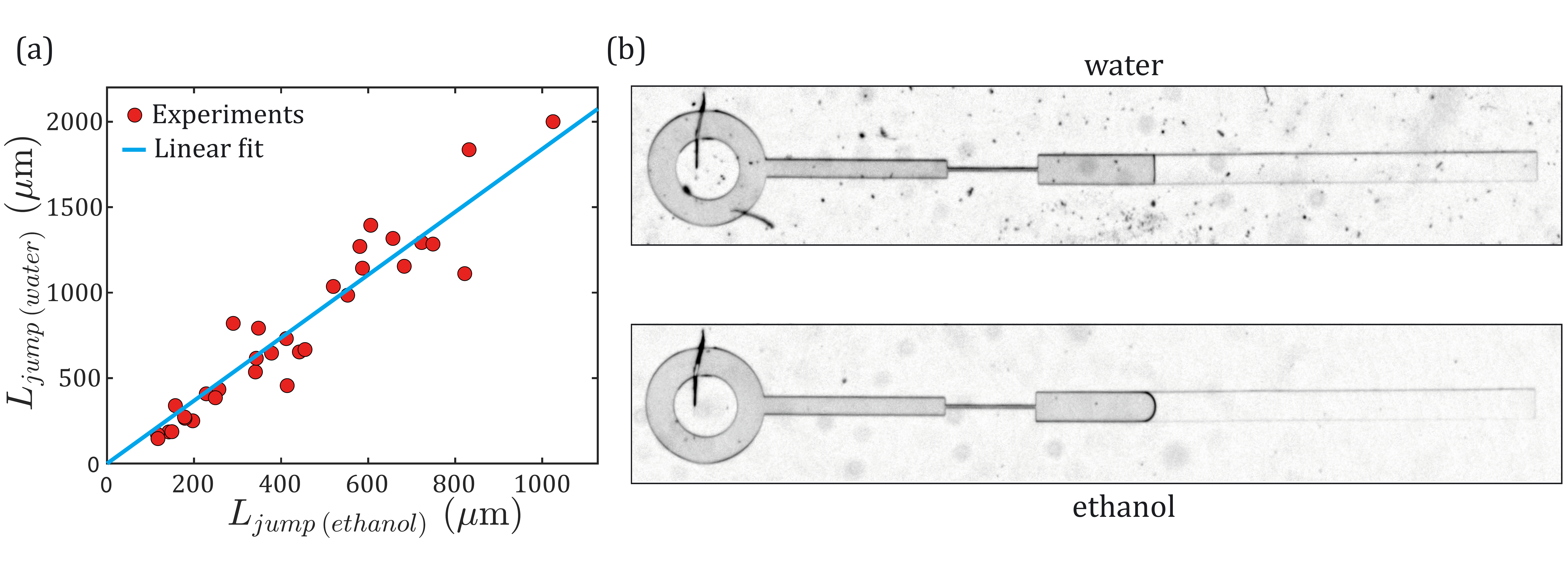}}% Images in 100% size
  \caption{(a) Plot of the length of the jump obtained with water $L_{\mathrm{jump}\:(\mathrm{water})}$ as a function of the length of the jump obtained with ethanol $L_{\mathrm{jump}\:(\mathrm{ethanol})}$ in channels with various geometrical parameters. The straight line corresponds to the best linear fit $L_{\mathrm{jump}\:(\mathrm{water})} = 1.84\:L_{\mathrm{jump}\:(\mathrm{ethanol})}$.  (b) Two photographs of the air embolism present in the exit channel ($w_{\mathrm{out}}=330$ $\mu$m, $h=53$ $\mu$m, $H=90$ $\mu$m), for both water (top) and ethanol (bottom). The interface curvature between air (at the left) and liquid (at the right) is markedly larger for ethanol than for water, due to a better wettability.}
\label{Fig:ethanol}
\end{figure}

The experiments were also carried out using ethanol as the liquid. Similar jumps were observed, characterised by a length $L_{\mathrm{jump}\:(\mathrm{ethanol})}$. When comparing $L_{\mathrm{jump}\:(\mathrm{ethanol})}$ with the jump length $L_{\mathrm{jump}\:(\mathrm{water})}$ obtained with water in the same geometry, we observe the jump is systematically shorter with ethanol than with water, as represented in figure \ref{Fig:ethanol}a. Experimental data are well adjusted by the linear relationship $L_{\mathrm{jump}\:(\mathrm{water})}=1.84\:L_{\mathrm{jump}\:(\mathrm{ethanol})}$.

This good linear correlation agrees with our scenario involving a capillary threshold, itself proportional to surface tension. However, the prefactor 1.84 is smaller than the surface tension ratio between water and ethanol, which equals 3.3 (taking 22~mN/m for the surface tension of ethanol). Two main reasons may explain this discrepancy. First, the maximal curvature of the meniscus is larger for ethanol than for water due to better wetting conditions, as shown on the two comparative pictures of figure \ref{Fig:ethanol}b; hence, the ratio of maximal Laplace pressure at the capillary threshold is smaller than the ratio of surface tension. Second, in our experiments, the surface tension of water might be smaller than that of pure water, due to pollution of the interface by the extraction of the uncrosslinked PDMS chains. Indeed, \citet{Hourlier-Fargette2017} showed that the surface tension of water droplets could drop to $52$ mN/m in presence of such a pollution.

\subsection{Volume conservation during the jump}
\label{part:volume}

\begin{figure}
  \centerline{\includegraphics[width=0.5\columnwidth]{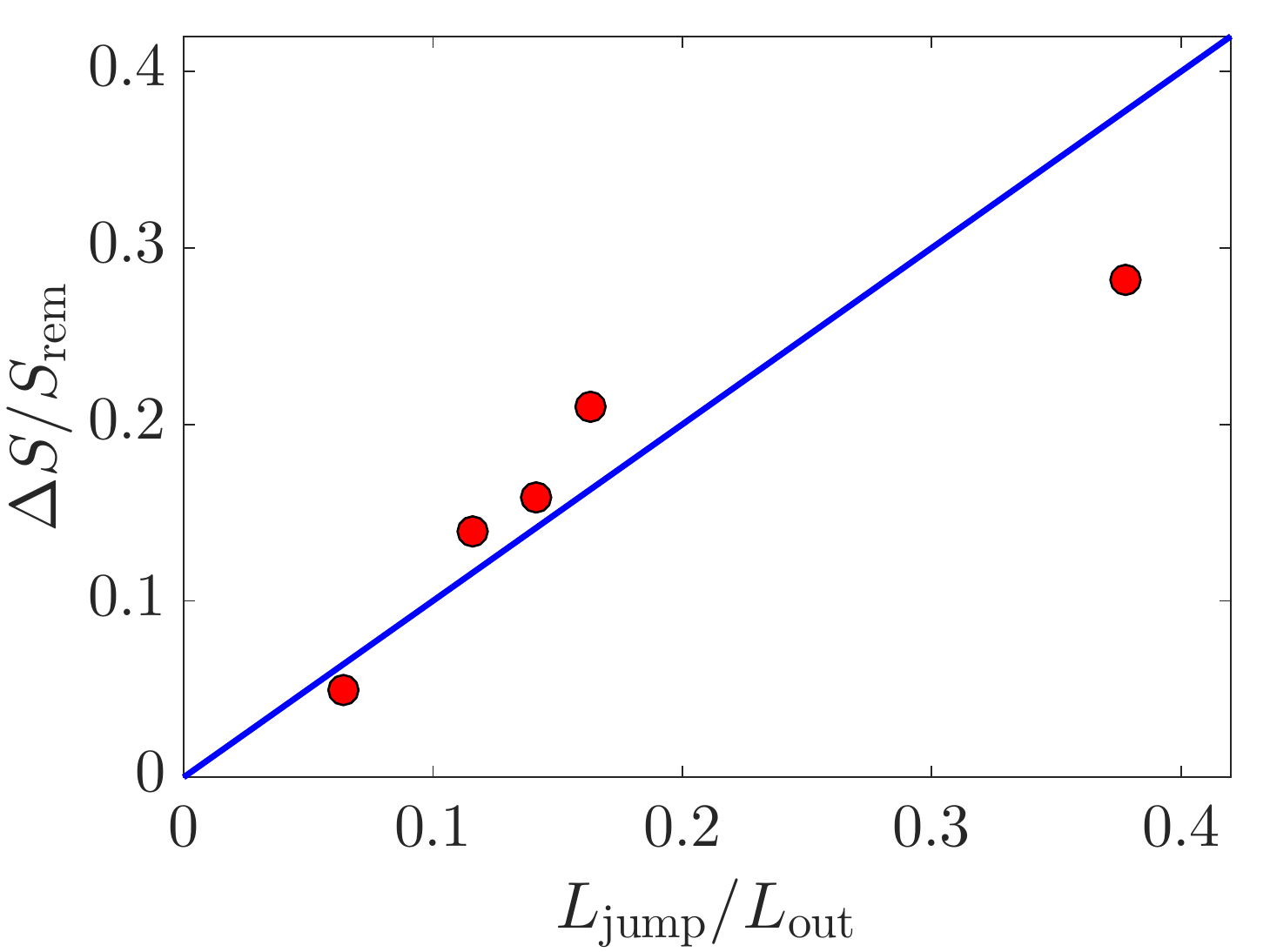}}% Images in 100% size
  \caption{Plot of the relative deformation $\Delta S/S_{\mathrm{rem}}$ observed by confocal imaging, compared to the relative amplitude of the jump $L_{\mathrm{jump}}/L_{\mathrm{out}}$ measured in a second round of experiments with macroscopic observations (red circles). The blue line represents the conservation of water volume during the jump (equation \ref{eq:conservation}).}
\label{Fig:Volume_conservation}
\end{figure}

The characteristic time scale of the jump, varying between 0.1 and 2~s depending on the experiments is systematically much smaller than the characteristic time scale of pervaporation $\tau_{\mathrm{perv}} = w_{\mathrm{out}}/|\dot{L}|$ (with $|\dot{L}|$ the interface speed just after the jump), which is itself of order $10^3$ to $10^4$~s. We can thus consider that the volume of liquid is conserved during the jump, which implies: $S_{\min} L_{\mathrm{out}} = S_{\mathrm{rem}} (L_{\mathrm{out}} - L_{\mathrm{jump}})$, with $S_{\min}$ the cross-sectional area just before the jump. Hence:
\begin{equation} \label{Eq:L_jump(DeltaS)}
    \frac{L_{\mathrm{jump}}}{L_{\mathrm{out}}} = \frac{\Delta S}{S_{\mathrm{rem}}} ,
    \label{eq:conservation}
\end{equation}
where $\Delta S = S_{\mathrm{rem}}-S_{\min}$. For a few experiments, we compared the confocal measurements of  relative deformation of the channel before and after the jump $\Delta S/S_{\mathrm{rem}}$ with the relative amplitude of the jump $L_{\mathrm{jump}}/L_{\mathrm{out}}$ measured from macroscopic observations realised with the same channels a few days after the confocal imaging experiments. Figure \ref{Fig:Volume_conservation} shows that the data are compatible with this scenario of volume conservation during the jump, within a certain margin of error due to the confocal imaging resolution and the relative reproducibility of $L_{\mathrm{jump}}$ discussed earlier. 

%Altogether, the experimental results provided in this section demonstrate that the abrupt propagation of the air embolism originates from the compliance of the channel and does not result from strong variations of the evapotranspiration rate during the embolization.

\section{Effect of geometrical parameters on jump length and arrest time}

The results shown in section \ref{Sec:description_dynamics} point with good confidence towards the following scenario. The meniscus gets pinned at the constriction corners for a certain time. During this arrest phase, the ongoing drying induces a decrease of the channel cross-section, hence a decrease of the water pressure, and bending of the upper wall. Due to Laplace pressure, this increases the in-plane curvature of the meniscus until a capillary threshold, where a sudden relaxation occurs: the cross-section recovers quickly its equilibrium value, and the meniscus jumps quickly forwards. In this new section, we reinforce the validity of the depicted mechanism by evidencing the influence of the geometrical parameters of the channel (length $L_{\mathrm{out}}$, width $w_{\mathrm{out}}$, membrane thickness $\delta$ and constriction width) on the length of the jump $L_{\mathrm{jump}}$ (sections \ref{part:l_out}, \ref{Sec:compliance} and \ref{part:w_p}) and the time of residence in the constriction $t_{\mathrm{res}}$ (section \ref{part:t_res}).

\subsection{The jump length $L_{\mathrm{jump}}$ is proportional to the channel length $L_{\mathrm{out}}$}
\label{part:l_out}

\begin{figure}
    \centering
    \begin{subfigure}[]
        {\includegraphics[width=0.45\textwidth]{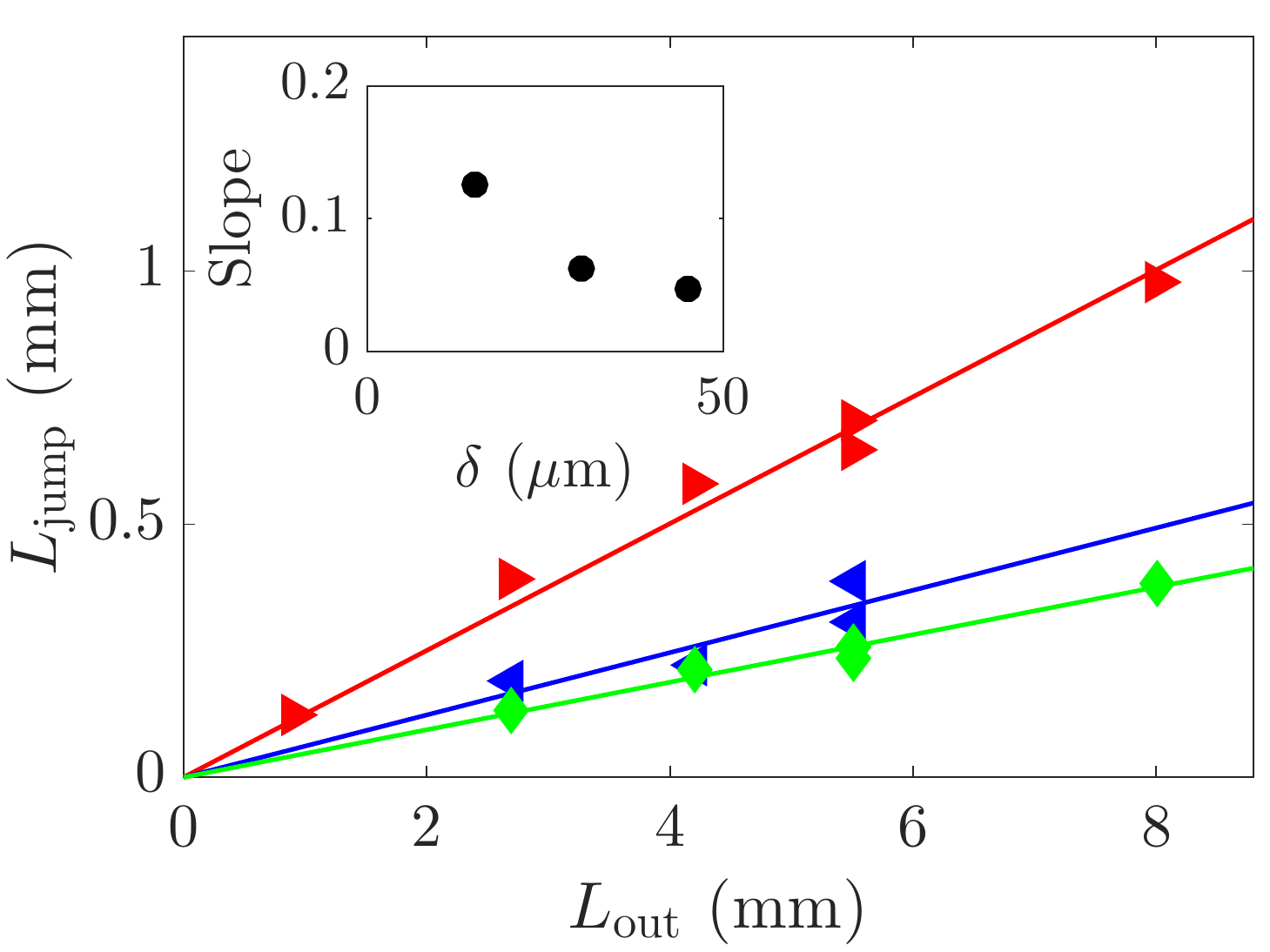}}
        %\caption{$h = 35~\mu$m, $H = 75~\mu$m.}
        %\label{Fig:L(t_tc)_h035um_H075um}
    \end{subfigure}
    ~ %add desired spacing between images, e. g. ~, \quad, \qquad, \hfill etc. 
      ~\begin{subfigure}[]
        {\includegraphics[width=0.45\textwidth]{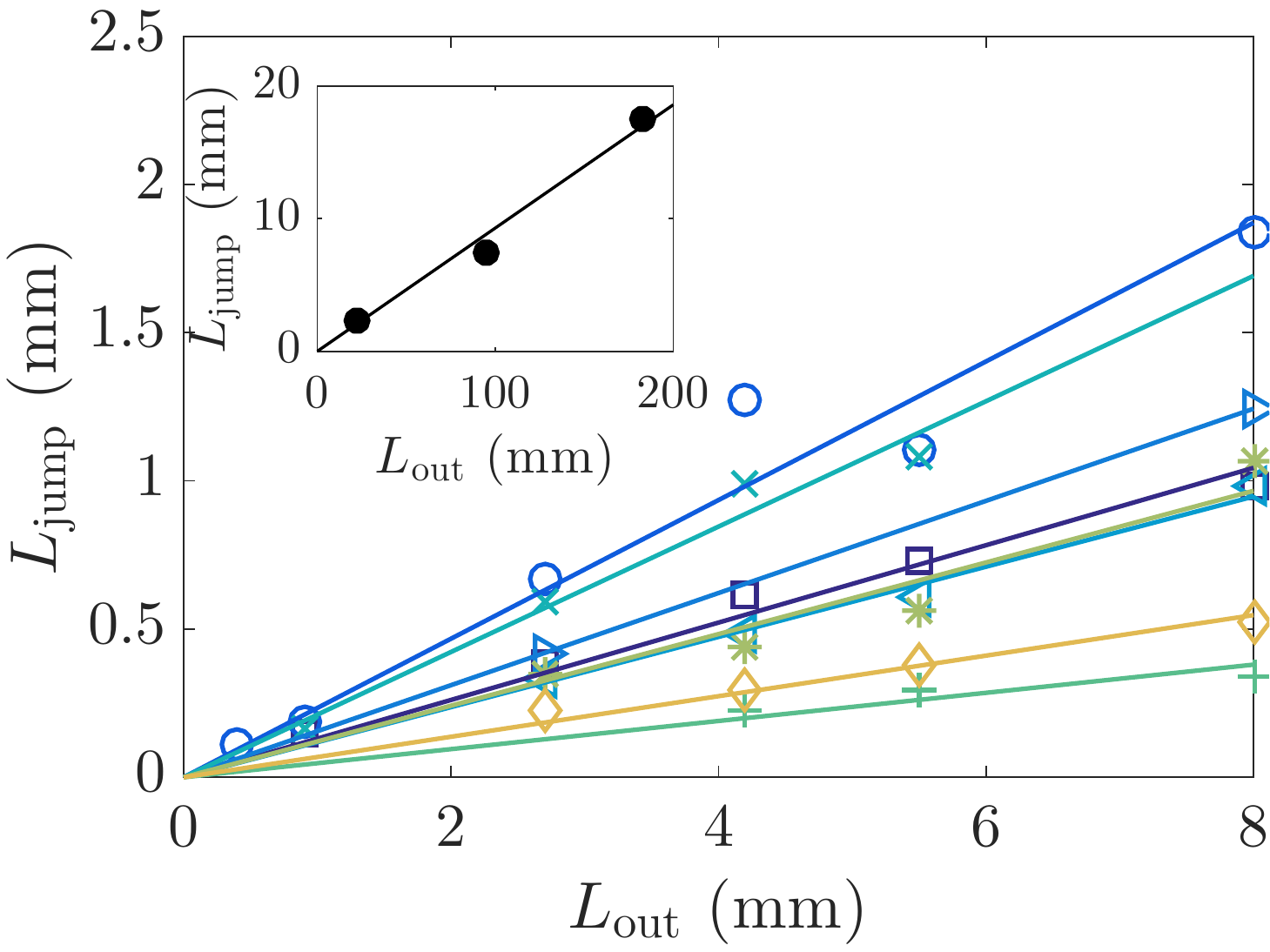}}
        %\caption{$h = 37~\mu$m, $H = 102~\mu$m.}
        %\label{Fig:L(t_tc)_h037um_H102um}
    \end{subfigure}
    \caption{(a) Plot of the jump length $L_{\mathrm{jump}}$ as a function of the length of the exit channel $L_{\mathrm{out}}$, at fixed values of the exit channel width $w_{\mathrm{out}} = 300~\mu$m and channel height $h = 65~\mu$m, and for three values of the upper wall thickness: $\delta = 15$ ($\blacktriangleright$), 30 ($\blacktriangleleft$) and 45~$\mu$m ($\blacklozenge$). Straight lines are the best linear fits of the three series of data, and the best fitting slope is plotted as a function of $\delta$ in the inset. (b) Plot of the jump length $L_{\mathrm{jump}}$ as a function of the length of the exit channel $L_{\mathrm{out}}$ in channels of width $w_{\mathrm{out}} = 300~\mu$m, for couples of values of $(h,H)$ equal (in $\mu$m) to: 53, 90 ($\square$); 44, 78 ($\circ$); 53, 79 ($\triangleright$); 65, 89 ($\triangleleft$); 44, 64 ($\times$); 100, 154 ($+$); 53, 85 ($*$); 53, 105 ($\lozenge$). Inset: Plot of the jump length $L_{\mathrm{jump}}$ as a function of the length of the exit channel $L_{\mathrm{out}}$ in long channels of width $w_{\mathrm{out}} = 350~\mu$m, for $h = 67~\mu$m and $H = 111~\mu$m.} \label{Fig:Ljump_vs_Lout}
\end{figure}

We first report measurements of the jump length as a function of the length the exit channel, for various experiments, in figure \ref{Fig:Ljump_vs_Lout}. It confirms the water volume conservation during the jump (section \ref{part:volume}), namely that the jump length is simply proportional to the length of the exit channel with an excellent correlation. Moreover, figure \ref{Fig:Ljump_vs_Lout}a shows that, at given exit channel width and channel height, the coefficient of proportionality decreases at increasing upper wall thickness. This proportionality also holds in much longer exit channels (inset of figure \ref{Fig:Ljump_vs_Lout}b). %its length $L_{\mathrm{out}}$, its width $w_{\mathrm{out}}$, its height $h$ and the PDMS thickness $H$ or, alternatively, the upper wall thickness $\delta = H - h$. [COMMENTER L'EFFET DE $w_p$]

%Starting from an experiment with parameters $L_{\mathrm{out}} = 5.5$~mm, $w_{\mathrm{out}} = 300~\mu$m, $h = 65~\mu$m and $H = 95~\mu$m, we vary one or two parameter(s) at a time, to isolate the effect of each geometrical parameter. First, figure \ref{Fig:Ljump_ref}a shows that the jump length is simply proportional to the length of the exit channel with an excellent correlation, with a coefficient of proportionality which decreases at increasing upper wall thickness. Second, figure \ref{Fig:Ljump_ref}b shows that the jump length increases at increasing width of the exit channel, the increase being more marked as the upper wall thickness increases. Third, figure \ref{Fig:Ljump_ref}c shows that the jump length decreases at increasing upper wall thickness.

The proportionality between the jump length and the length of the exit channel has been interpreted in section \ref{part:volume}. To make the discussion more complete, it is worth noticing that the quantity $\Delta S$ appearing in (\ref{Eq:L_jump(DeltaS)}) depends on the channel compliance and the capillary threshold. Neither of these two quantities depend on $L_{\mathrm{out}}$: the compliance depends on $w_{\mathrm{out}}$, $h$ and $H$, while the capillary threshold depends on $h$ and $w_{\mathrm{p}}$. This further explains the proportionality between the jump length and the length of the exit channel.

\subsection{Effect of the channel compliance on the jump length} \label{Sec:compliance}

\begin{figure}
    \centering
    \begin{subfigure}[]
        {\includegraphics[width=0.45\textwidth]{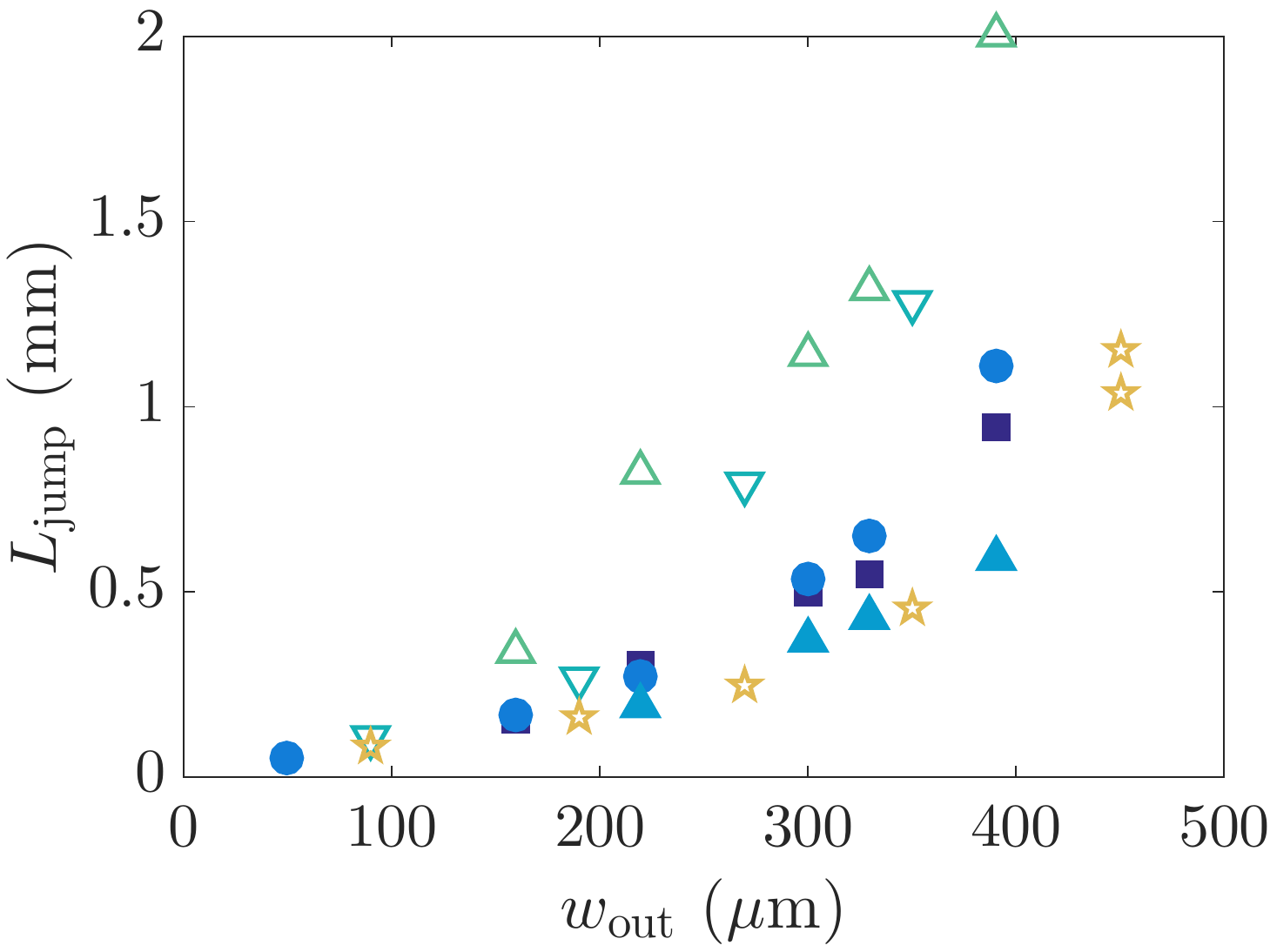}}
        %\caption{$h = 35~\mu$m, $H = 75~\mu$m.}
        %\label{Fig:L(t_tc)_h035um_H075um}
    \end{subfigure}
    ~ %add desired spacing between images, e. g. ~, \quad, \qquad, \hfill etc. 
      ~\begin{subfigure}[]
        {\includegraphics[width=0.45\textwidth]{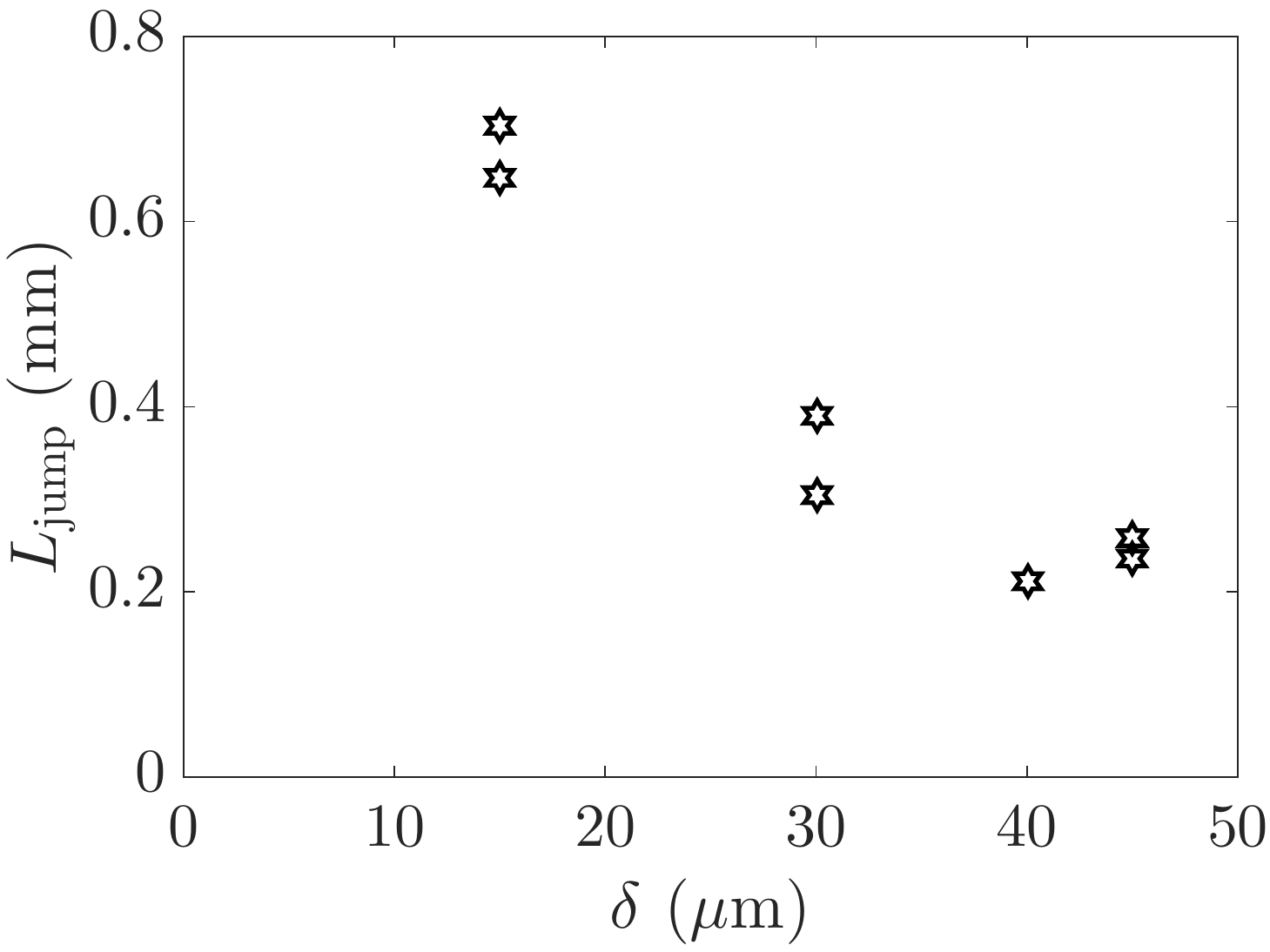}}
        %\caption{$h = 37~\mu$m, $H = 102~\mu$m.}
        %\label{Fig:L(t_tc)_h037um_H102um}
    \end{subfigure}
    \caption{%(a) Plot of the jump length $L_{\mathrm{jump}}$ as a function of the length of the exit channel $L_{\mathrm{out}}$ in channels of width $w_{\mathrm{out}} = 300~\mu$m, for couples of values of $(h,H)$ equal (in $\mu$m) to: 53, 90 ($\square$); 44, 78 ($\circ$); 53, 79 ($\triangleright$); 65, 89 ($\triangleleft$); 44, 64 ($\times$); 100, 154 ($+$); 53, 85 ($*$); 53, 105 ($\lozenge$). Inset: Plot of the jump length $L_{\mathrm{jump}}$ as a function of the length of the exit channel $L_{\mathrm{out}}$ in long channels of width $w_{\mathrm{out}} = 350~\mu$m, for $h = 67~\mu$m and $H = 111~\mu$m.
    (a) Plot of the jump length $L_{\mathrm{jump}}$ as a function of the width of the exit channel $w_{\mathrm{out}}$ in channels of length $L_{\mathrm{out}} = 5.5$~mm, for a fixed channel height $h = 53~\mu$m and PDMS thickness $H = 85$ ($\blacksquare$), 90 ($\bullet$) and 105~$\mu$m ($\blacktriangle$), and a fixed upper wall thickness $\delta = 34.5 \pm 1~\mu$m and channel height $h = 35$ ($\triangledown$), 44 ($\triangle$), 53 ($\bullet$) and 60~$\mu$m (pentagram symbols). (b) Plot of the jump length $L_{\mathrm{jump}}$ as a function of the upper wall thickness $\delta$, at fixed values of the length of the exit channel $L_{\mathrm{out}} = 5.5~$mm, width of the exit channel width $w_{\mathrm{out}} = 300~\mu$m, and channel height $h = 65~\mu$m.} \label{Fig:Ljump_vs_wout_delta}
\end{figure}

In order to quantify the effect of the channel compliance, we measured the jump length as a function of the width and the height of the channel and of the upper wall thickness, for various experiments where the length of the exit channel is fixed. We show the influence of the channel width $w_{\mathrm{out}}$ in figure \ref{Fig:Ljump_vs_wout_delta}a, and that of the upper wall thickness in figure \ref{Fig:Ljump_vs_wout_delta}b. Figure \ref{Fig:Ljump_vs_wout_delta}a shows that for each value of $(h,H)$, the jump length increases at increasing channel width, with a convex trend. Figure \ref{Fig:Ljump_vs_wout_delta}b shows that at given channel width and height, the jump length decreases at increasing upper wall thickness.

\begin{figure}
  \centerline{\includegraphics[width=0.5\columnwidth]{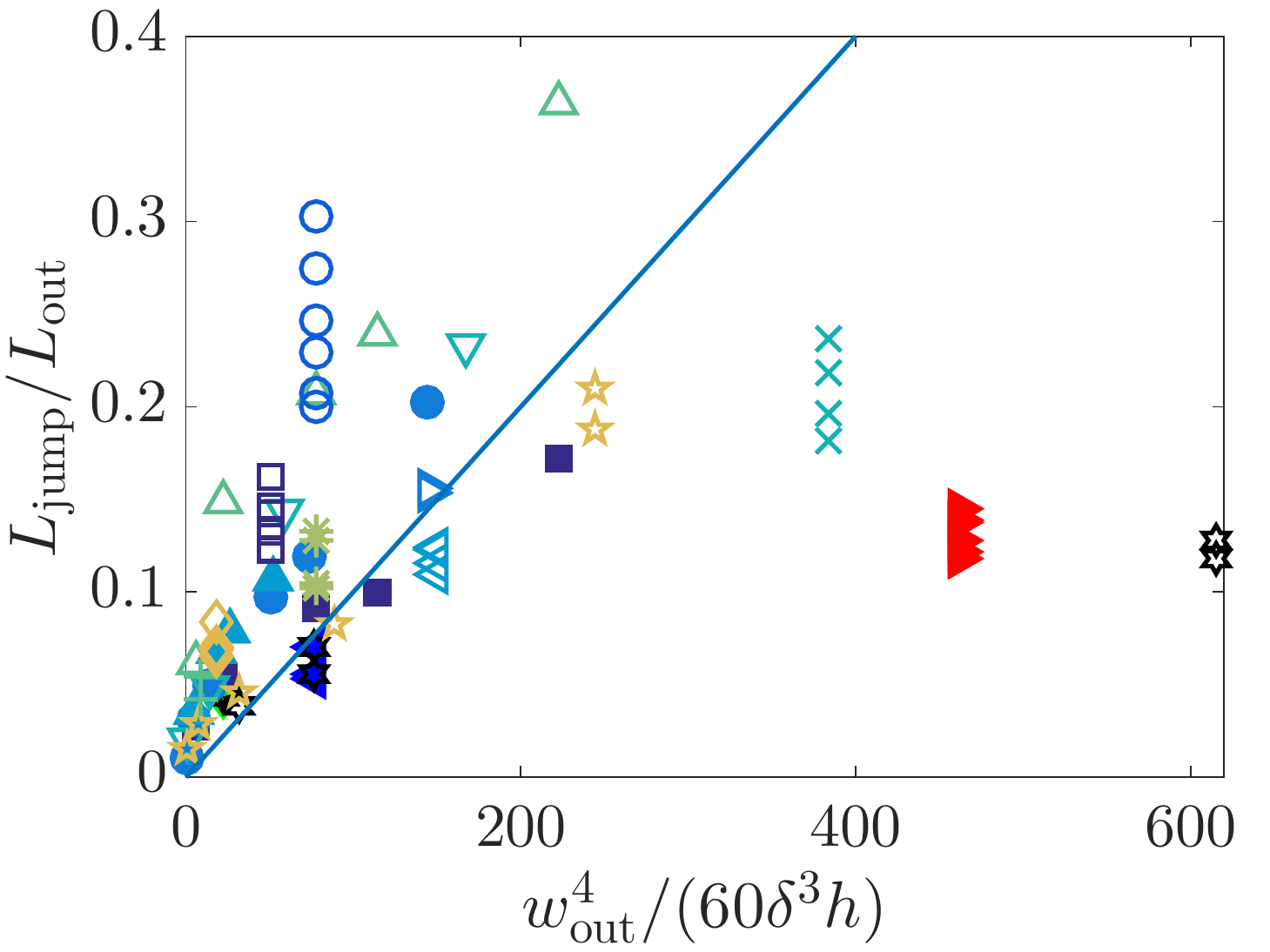}}% Images in 100% size
  \caption{Plot of the ratio $L_{\mathrm{jump}}/L_{\mathrm{out}}$ as a function of $w_{\mathrm{out}}^4/(60E\delta^3 h)$. The straight line is the linear law of slope $10^{-3}$. Different symbols correspond to the various experiments shown in figures \ref{Fig:Ljump_vs_Lout} and \ref{Fig:Ljump_vs_wout_delta}.}
\label{Fig:deformabilite}
\end{figure}

The dependence of the jump length on the channel width and upper wall thickness may be discussed in relation with the thin beam approach which was used in section \ref{Sec:description_dynamics} to describe the deflection of the upper wall. From (\ref{Eq:deflection}) and (\ref{Eq:L_jump(DeltaS)}), it is immediate to compute
$\Delta S$:
$$ \Delta S = \int_{-w_{\mathrm{out}}/2}^{w_{\mathrm{out}}/2} \zeta \, \mathrm{d}y = \frac{w_{\mathrm{out}}^5 \Delta P}{720B} , $$
whence, from (\ref{Eq:L_jump(DeltaS)}), the prediction:
\begin{equation} \label{Eq:prediction_deformabilite}
    \frac{L_{\mathrm{jump}}}{L_{\mathrm{out}}} \simeq \frac{w_{\mathrm{out}}^4}{60\delta^3 h} \frac{(1 - \nu^2) \Delta P}{E} ,
    \label{eq:jump_prediction}
\end{equation}
neglecting the difference between $S_0$ and $S_{\mathrm{rem}}$ (see the mention to remanent pressure in section \ref{part:confocal}). Hence, the jump length is predicted to be an increasing function of the channel width and a decreasing function of the upper wall thickness, in qualitative agreement with figure \ref{Fig:Ljump_vs_wout_delta}. The relation (\ref{Eq:prediction_deformabilite}) also shows that the ratio $L_{\mathrm{jump}}/L_{\mathrm{out}}$ should be proportional to the product of the geometrical dimensionless parameter $w_{\mathrm{out}}^4/(60E\delta^3 h)$, which quantifies the ability of the channel to deform under a given load, and of $(1 - \nu^2)\Delta P/E$. The latter quantity was shown to equal $10^{-3}$ in section \ref{part:elastcapillarity}. Since the constriction width $w_{\mathrm{p}}$ remains fixed for this data set, we expect the capillary threshold setting the pressure difference $\Delta P$ not to vary much between all those experiments. Hence, we plot in figure \ref{Fig:deformabilite} the ratio $L_{\mathrm{jump}}/L_{\mathrm{out}}$ as a function of $w_{\mathrm{out}}^4/(60E\delta^3 h)$, with the straight line of slope $10^{-3}$, gathering all data from figures \ref{Fig:Ljump_vs_Lout} and \ref{Fig:Ljump_vs_wout_delta}. It shows that (\ref{Eq:prediction_deformabilite}) gives the good trend and the good order of magnitude for the jump length as a function of parameters setting the channel compliance, but with a large dispersion. We can evoke several reasons for the limited rescaling of the data: (i) the ratio $\delta/w_{\mathrm{out}}$ is not much larger than one, hence the thin beam theory is only approximate in our experimental conditions. (ii) The clamped end conditions are also approximate, because the bulk of PDMS to which the upper wall is attached is of course not infinitely rigid and may deform. (iii) The capillary threshold may also vary, because it is sensitive to the details of geometry of the constriction exit. Indeed, because of the limited precision of the microfabrication process, the exit corners are not sharp but blunt, with a radius of curvature which may differ between different channels.

It is worth mentioning that our discussion actually concerns the hydraulic capacitance of the channel, a concept used frequently in hydraulics of plants \citep{Brodribb2005} and in microfluidics \citep{Bruus2008}. The capacitance links the incoming flux of fluid $Q_f$ to the variation of pressure with $Q_f = C_h \mathrm{d}\Delta P/\mathrm{d}t$. Here the capacitance is due to elastic response  of the channel volume, and from (\ref{eq:jump_prediction}) we derive (introducing $V_{\mathrm{channel}} = hw_{\mathrm{out}} L_{\mathrm{out}}$ the volume of the channel): $$C_h=\frac{\mathrm{d}V_{\mathrm{channel}}}{\mathrm{d}\Delta P}=\frac{L_{\mathrm{out}} w_{\mathrm{out}}^5}{60\delta^3 } \frac{1 - \nu^2}{E},$$ a quantity useful to characterise the change of channel volume when pressure varies.

% COMMENTER LES D\'EPENDANCES AVEC LES DIFF\'ERENTS PARAM\`ETRES G\'EOM\'ETRIQUES. PROPORTIONNALIT\'E AVEC $L_{\mathrm{out}}$. D\'EPENDANCES AVEC $w_{\mathrm{out}}$ et $\delta$ RAISONNABLES, MAIS MOINS FORTES QU'ATTENDU AVEC UN MOD\`ELE DE PLAQUE MINCE, PEUT-\^ETRE PARCE QU'ON S'\'ECARTE D'UN TEL MOD\`ELE. PEUT-\^ETRE AJUSTER L'ENSEMBLE DES DONN\'EES $L_{\mathrm{jump}}$ AVEC CE MOD\`ELE DE PLAQUE MINCE ET DISCUTER LES D\'EVIATIONS~?

\subsection{Influence of the constriction width $w_{\mathrm{p}}$.}
\label{part:w_p}

\begin{figure}
  \centerline{\includegraphics[width=0.5\columnwidth]{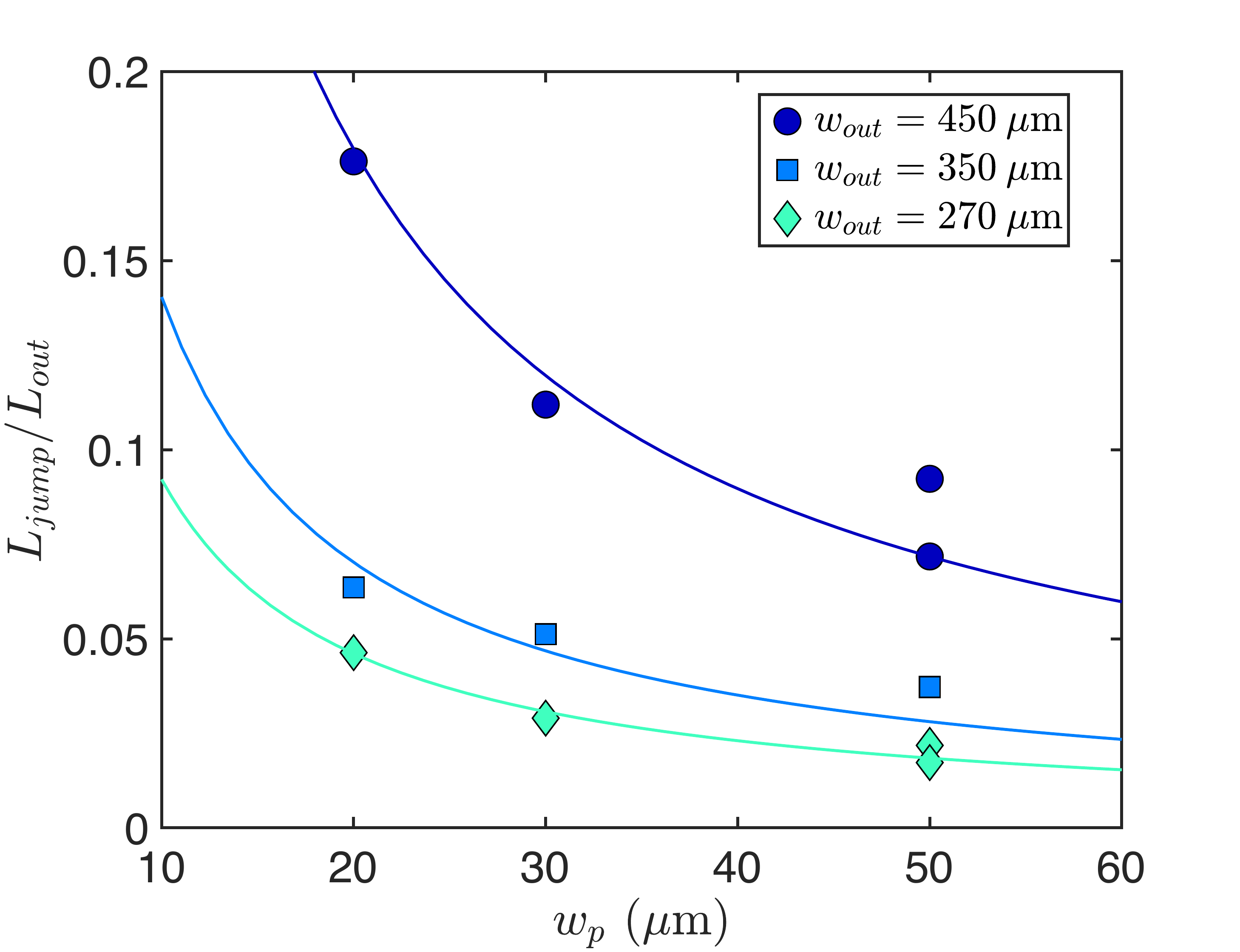}}% Images in 100% size
  \caption{Plot of the ratio $L_{\mathrm{jump}}/L_{\mathrm{out}}$ as a function of the constriction width $w_{\mathrm{p}}$ for three different channel widths $w_{\mathrm{out}}$ (270, 350 and 450 $\mu$m), with $h=67\:\mu$m, $H=95\:\mu$m and $L_{\mathrm{out}}=11$ or $22$ mm. The straight lines correspond to the best inverse fits: $L_{\mathrm{jump}}/L_{\mathrm{out}} = a\cdot w_{\mathrm{p}}^{-1} $, with $a$ a fit parameter.}
\label{Fig:constrictionwidth}
\end{figure}

Using channels of various constriction widths $w_{\mathrm{p}}$, we now discuss the influence of the minimal radius of curvature of the interface past the constriction. The mechanism described earlier accounts for a deformation of the channel induced by the Laplace depression in the liquid, itself generated by the strong curvature of the interface in the constriction. It is thereby expected that the jump would be larger for stronger curvature, i.e. for smaller $w_{\mathrm{p}}$. Figure \ref{Fig:constrictionwidth} shows three different set of measurements, for three different channel widths $w_{\mathrm{out}}$. It shows that the length of the jump $L_{\mathrm{jump}}$ systematically decreases as the width of the constriction increases, in agreement with the physical mechanism proposed in last section. Although the minimal radius of curvature of the interface is not exactly $w_{\mathrm{p}}/2$ due to the finite angle of the constriction exit (figure \ref{Fig:meniscus_curvature}), data are compatible with inverse power laws (solid lines in figure \ref{Fig:constrictionwidth}) deriving from (\ref{eq:jump_prediction}) with $\Delta P \approx -\gamma/w_{\mathrm{p}}$.

\subsection{Residence time within the constriction $t_{\mathrm{res}}$}
\label{part:t_res}

Another experimental quantity of interest is the total residence time of the meniscus inside the constriction $t_{\mathrm{res}}$ (shown in figure \ref{Fig:L_t_ref}b), which is the counterpart of the time of retention of an air embolism in a tree before propagation in the neighbouring vessel. During this time, the volume of water which leaves the channel by pervaporation is the sum of the volume $hw_{\mathrm{p}} L_{\mathrm{p}}$ of the constriction and of the volume $L_{\mathrm{out}} \Delta S = L_{\mathrm{jump}} S_{\mathrm{rem}}$. The latter contribution comes from our interpretation of the jump length in terms of water conservation leading to the relation (\ref{Eq:L_jump(DeltaS)}). Hence, the residence time inside the constriction should equal $t^* = (L_{\mathrm{jump}} S_{\mathrm{rem}} + hw_{\mathrm{p}} L_{\mathrm{p}})/Q$, with $Q$ the water flux leaving the channel. For simplicity, we neglect the flux from the water-filled part of the constriction. Therefore, according to the predictions of \citet{Dollet2019} recalled in section \ref{part:smooth}, the water flux equals $Q = q_\ell L_{\mathrm{out}} + Q_g$, with $q_\ell$ the diffusion flux per unit length along the exit channel, and $Q_g$ the evaporation flux at the meniscus. We thus get the following prediction:
\begin{equation} \label{Eq:prediction_t*}
    t^* = \frac{L_{\mathrm{jump}} S_{\mathrm{rem}} + hw_{\mathrm{p}} L_{\mathrm{p}}}{q_\ell L_{\mathrm{out}} + Q_g} ,
\end{equation}
with $q_\ell$ given by (\ref{Eq:ql}), and where, from (\ref{Eq:Qg}), $Q_g = \sqrt{\alpha D_a D_P hw_{\mathrm{p}} \tilde{q}_\ell(w_{\mathrm{p}})}$. The latter flux $Q_g$ depends on the dimensions of the channel upstream the meniscus, whence the dependence on the constriction width $w_{\mathrm{p}}$, and not of the exit channel, in the factor $\tilde{q}_\ell$.

\begin{figure}
  \centerline{\includegraphics[width=0.5\columnwidth]{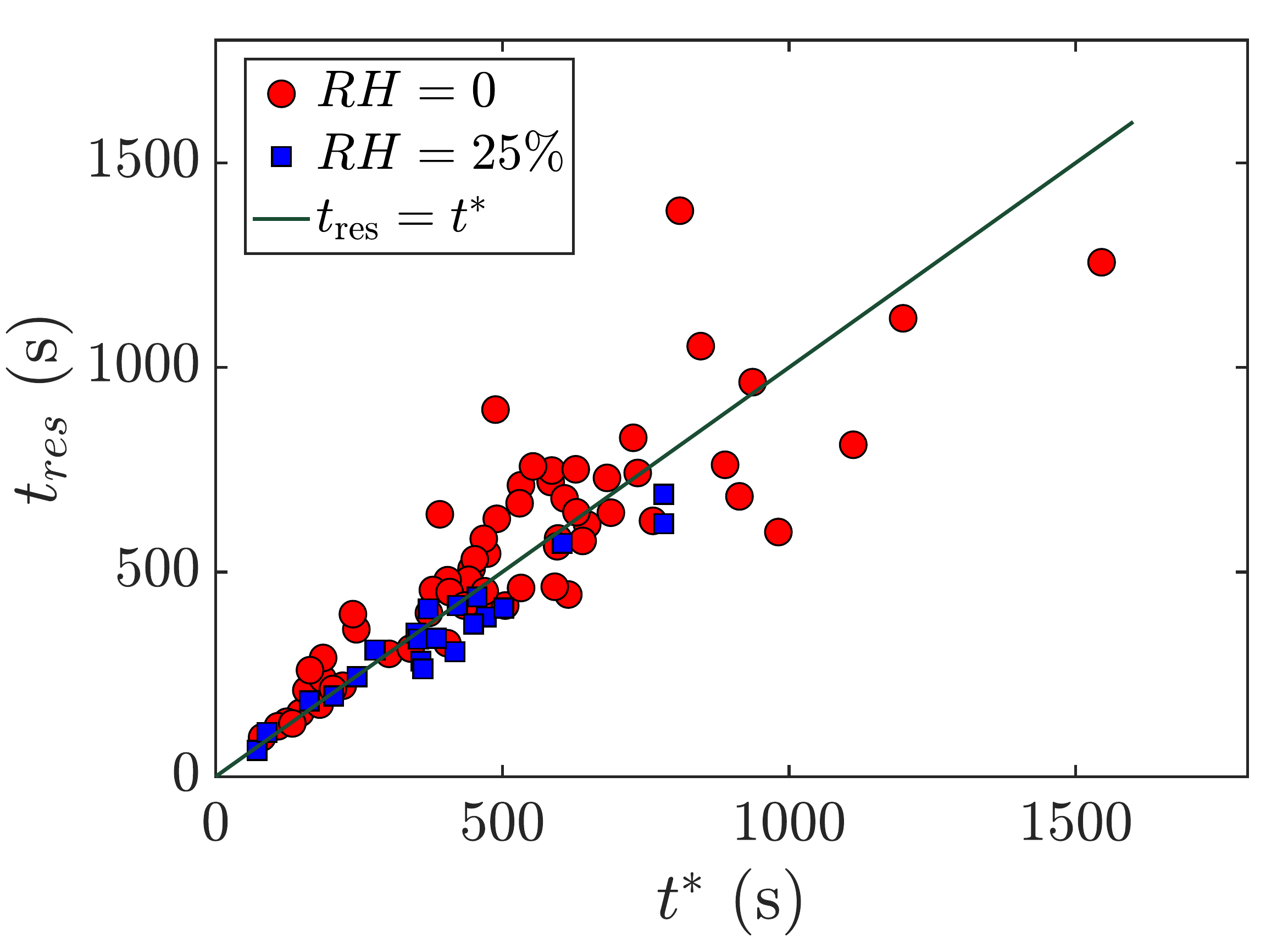}}% Images in 100% size
  \caption{Plot of the residence time of the meniscus inside the constriction $t_{\mathrm{res}}$ as a function of the time $t^*$ evaluated from Equation (\ref{Eq:prediction_t*}) (taking $S_{\mathrm{rem}} = hw_{\mathrm{out}}$), the value $D_P = 5.5 \times 10^{-10}$~m$^2$/s has been taken for the diffusivity of PDMS in water appearing in (\ref{Eq:ql}) and (\ref{Eq:Qg}). Data correspond to different relative humidity: $RH=0\:\%$ (red circles) and $RH=25\:\%$ (blue squares). The straight line corresponds to the equality $t_{\mathrm{res}} = t^*$.}
\label{Fig:temps_residence}
\end{figure}

In figure \ref{Fig:temps_residence}, we plot the residence time inside the constriction as a function of $t^*$ calculated using (\ref{Eq:prediction_t*}), where $D_P$ is adjusted so that the condition $t_{\mathrm{res}} = t^*$ is best fulfilled. We thus find $D_P = 5.5 \times 10^{-10}$~m$^2$/s. With this fitting value, figure \ref{Fig:temps_residence} shows indeed a good correspondence between the two times $t_{\mathrm{res}}$ and $t^*$. Furthermore, the order of magnitude of the aforementioned value of $D_P$ agrees with the value $10^{-9}$~m$^2$/s reported in the literature \citep{Watson1996}. Notice that none of the physicochemical parameters $D_P$, $\bar{c}_P^{\mathrm{sat}}$ and $\alpha$ encountered in the expressions (\ref{Eq:ql_tilde}) and (\ref{Eq:Qg}) of the fluxes is known with great precision. Hence, the fact that $t_{\mathrm{res}}$ and $t^*$ are found to be equal with reasonable values of these physicochemical parameters tends to confirm our interpretation of the residence time as the time required to pervaporate the water until the capillary threshold is reached.

\section{Relaxation dynamics following the jump}
\label{part:dynamics}
\subsection{Description of the dynamics}

\begin{figure}
  \centerline{\includegraphics[width=0.5\columnwidth]{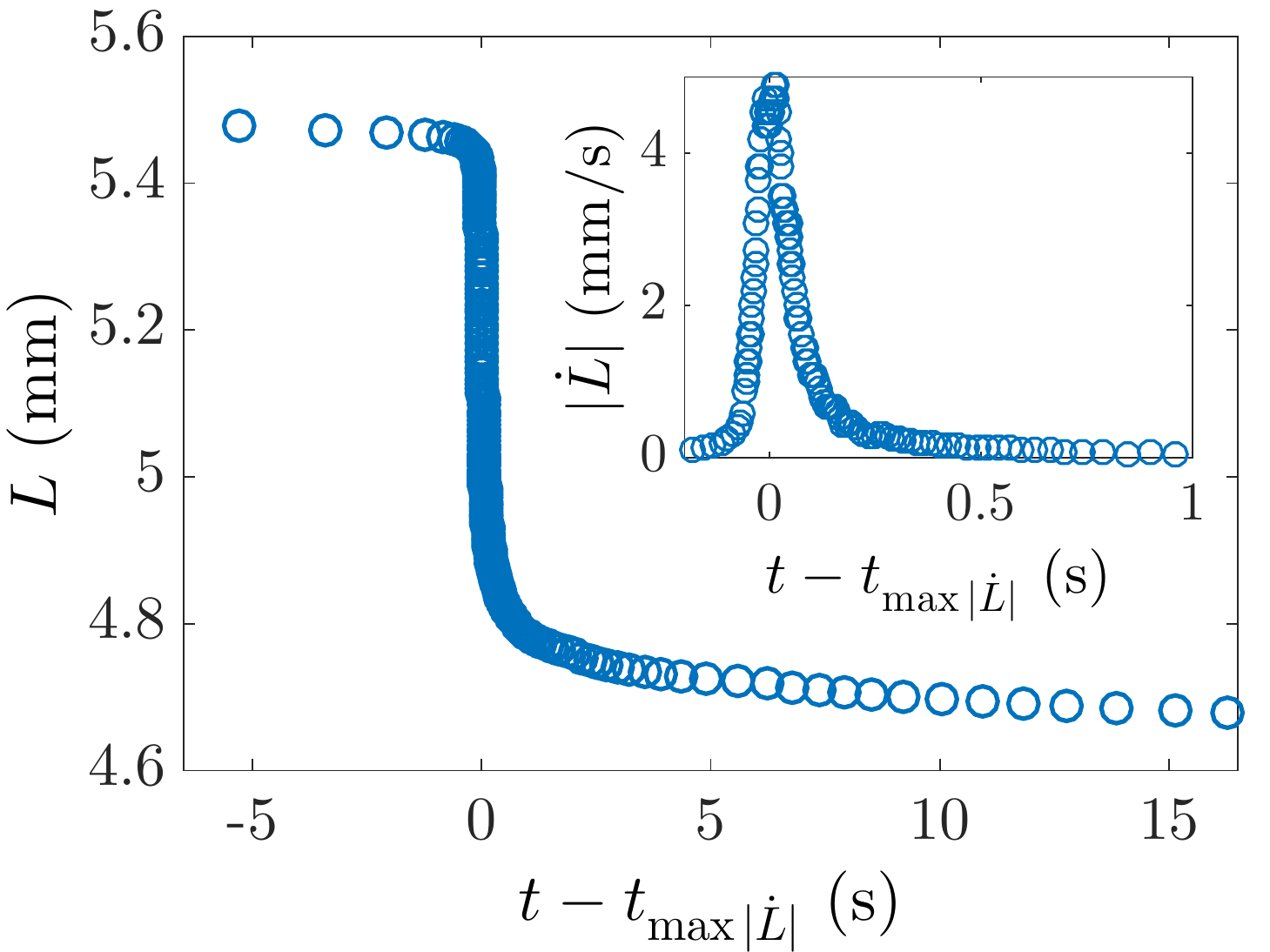}}% Images in 100% size
  \caption{Plot of the water length $L$ as a function of time $t - t_{\max|\dot{L}|}$ in a channel of geometrical parameters $L_{\mathrm{out}} = 5.5$~mm, $L_{\mathrm{p}} = 1$~mm, $w_{\mathrm{out}} = 390~\mu$m, $w_{\mathrm{p}} = 30~\mu$m, $h = 65~\mu$m and $H = 95~\mu$m. Inset: plot of the corresponding meniscus velocity $|\dot{L}|$ as a function of time in the fastest phase. The time $t_{\max|\dot{L}|}$ is the instant where the velocity reaches its maximal value.}
\label{Fig:L_v_rapide}
\end{figure}

We also studied the dynamics of the meniscus during the jump, by repeating the reference experiment studied in details in section \ref{Sec:description_dynamics}, using a high-speed camera. We supply as Supplementary Material a movie showing the jump and the subsequent relaxation. The water length is plotted as a function of time in figure \ref{Fig:L_v_rapide}, and an estimation of the instantaneous velocity of the meniscus $|\dot{L}|$ during the fastest phase is plotted as an inset of figure \ref{Fig:L_v_rapide}. This figure shows that the acceleration at the beginning of the jump phase is larger than the deceleration at the end of this phase. The maximal velocity is 5~mm/s, which is three orders of magnitude larger than the typical speeds of the meniscus in the entrance channel and in the exit channel after the jump (figure \ref{Fig:L_t_ref}a). This experiment also confirms the good repeatibility of the jump length.

\subsection{Physical discussion} \label{Sec:physical_interpretation_relaxation}

First, it is important to note that the dynamics of the jump (characteristic time scale smaller than $1$ s) is much faster than the pervaporation dynamics (characteristic time scale of about $1000$ s). This confirms that we can neglect the mass loss by pervaporation in the analysis, and consider a constant water volume in the channel throughout the process of the jump.

It is interesting to compare the relaxation dynamics observed in figure \ref{Fig:L_v_rapide} to the recent theoretical work by \citet{Martinez-Calvo2020}, who studied start-up flows in shallow deformable channels by imposing an initial jump in pressure or in flow rate, and predicted the subsequent dynamics. The corresponding transient flow followed unsteady lubrication theory, with negligible liquid inertia, and the upper wall was modelled by a thin plate under pure bending with clamped end conditions. Our experimental conditions are suited for the comparison. First, we proved that the channel upper wall behaves indeed as a thin plate bent by the pressure difference between the liquid and the outer atmosphere (figure \ref{Fig:deflection}). Second, our channel geometry reasonably fulfills the conditions $h \ll w_{\mathrm{out}} \ll L_{\mathrm{out}}$ assumed by \citet{Martinez-Calvo2020}. Third, liquid inertia is indeed negligible, as can be checked by evaluating the dimensionless parameter $\varepsilon\mathrm{Re} = \rho h^4 p_c/(12\eta^2 L_{\mathrm{out}}^2)$, where $p_c$ is the initial pressure jump. We have shown in section \ref{Sec:description_dynamics} that $p_c \approx \gamma/w_{\mathrm{p}} = 2$~kPa, which leads to the following numerical estimate: $\varepsilon\mathrm{Re} = 10^{-1}$, showing that inertia remains a secondary correction and that lubrication theory holds in good approximation. Fourth, the inertia of the top wall can be safely neglected; \citet{Martinez-Calvo2020} showed that the corresponding condition is that the dimensionless parameter defined by $\tilde{\gamma} = 3600\rho_s \delta h^6 B/(\eta^2 L_{\mathrm{out}}^2 w_{\mathrm{out}}^4)$, where $\rho_s = 1.0 \times 10^3$~kg/m$^3$ is the density of the solid, must be much smaller than 1. Numerical estimation with our parameters leads to $\tilde{\gamma} = 7 \times 10^{-8}$.

\begin{figure}
  \centerline{\includegraphics[width=0.5\columnwidth]{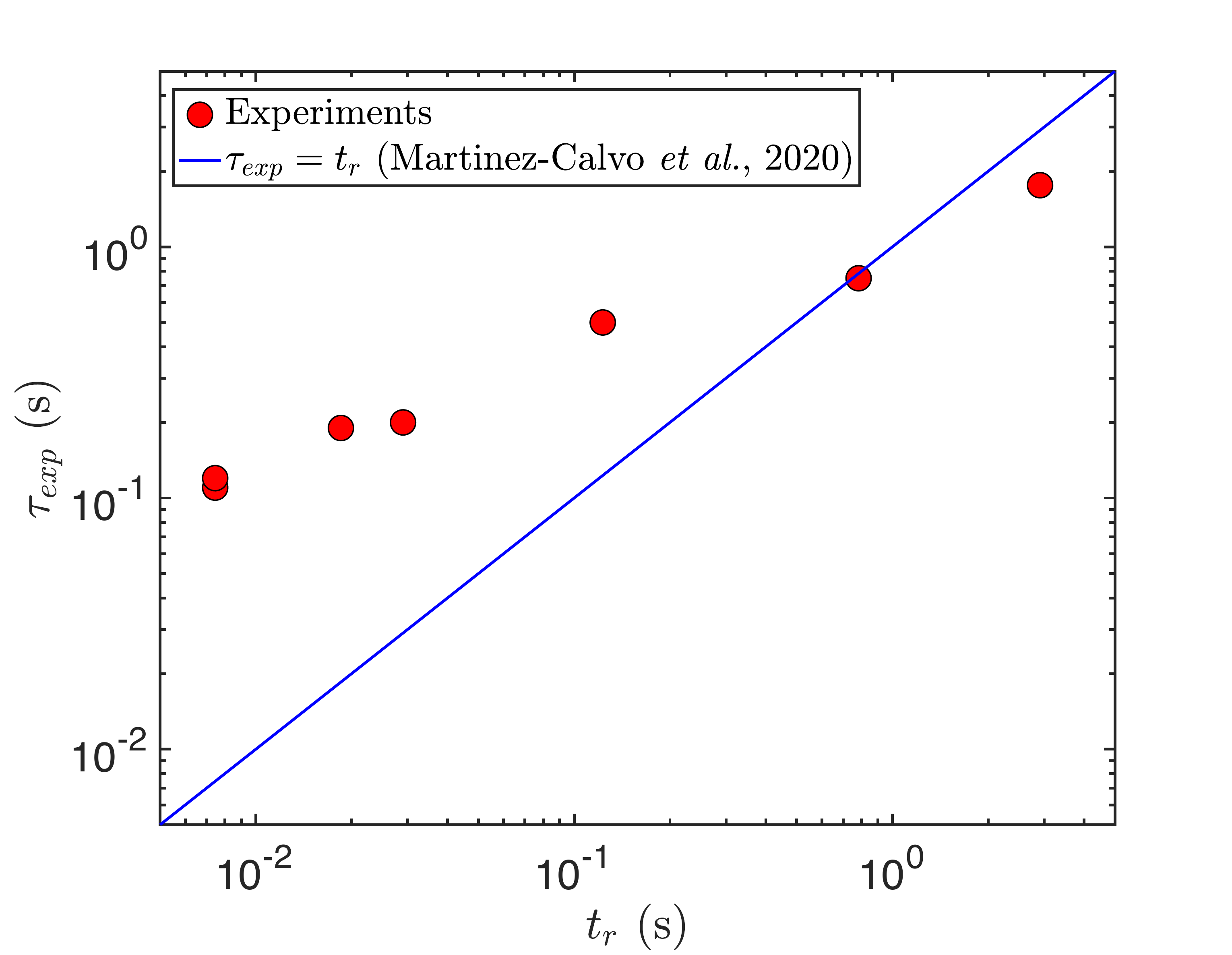}}% Images in 100% size
  \caption{Plot of the characteristic jump time $\tau_{\mathrm{exp}}$ measured experimentally as a function of the elasto-viscous characteristic time $t_r$ derived by \citet{Martinez-Calvo2020} (Equation \ref{eq:MC}). $\tau_{\mathrm{exp}}$ is obtained by an exponential fit of the $L(t)$ plots. Experiments were realized with the following combination of geometric parameters [$h$; $H$; $w_{\mathrm{out}}$; $L_{\mathrm{out}}$; $w_{\mathrm{p}}$] (in $\mathrm{\mu}$m) ordered by increasing value of the measured $\tau_{\mathrm{exp}}$: [11; 56; 100; 8000; 5], [60; 96; 450; 5500; 30], [67; 111; 450; 11000; 20], [67; 111; 450; 22600; 30], [67; 111; 350; 94400; 30], and [67; 111; 350; 182000; 30].}
\label{Fig:characteristic_time}
\end{figure}

Within such hypotheses, in the model of \citet{Martinez-Calvo2020}, the pressure in the channel relaxes following a nonlinear diffusion dynamics, the nonlinearity coming from the dependence of the hydrodynamic resistance on the channel cross-section. In our study, by virtue of the kinematic boundary condition and of the lubrication equations, the meniscus velocity is proportional to the pressure gradient of the liquid in contact. Hence, the temporal dynamics of the meniscus velocity can be compared to that of the pressure in the model of \citet{Martinez-Calvo2020}. In this comparison, the key quantity to estimate is the relaxation time of the diffusion dynamics. After a self-similar short-time regime, the long-time regime is an exponential relaxation with a characteristic time:
\begin{equation} \label{Eq:relaxation_time}
t_r = \frac{\eta w_{\mathrm{out}}^4 L_{\mathrm{out}}^2}{60h^3 B} ,
\label{eq:MC}
\end{equation}
see equation (4.5d) in \citet{Martinez-Calvo2020}. The numerical evaluation of this characteristic time is $t_r = 7$~ms, which is way too short to capture the relaxation dynamics shown in our experiments, which is of the order of $\tau_{\mathrm{exp}}\simeq0.2$ s (figure \ref{Fig:L_v_rapide}). However, experiments carried out in extremely long channels ($L_{\mathrm{out}}>10$ cm), and leading to remarkably long jumps ($L_{\mathrm{jump}}>1$ cm), exhibit characteristic times $\tau_{\mathrm{exp}}$ comparable to $t_r$ (see figure \ref{Fig:characteristic_time}). Note that these experimental characteristic times $\tau_{\mathrm{exp}}$ were obtaining after fitting the jump portion of the $L(t)$ curve by a common exponential function $L(t) = L_{\mathrm{out}} - b_1 [1 - \mathrm{e}^{-\left(t-t_{\mathrm{jump}}\right)/\tau_{\mathrm{exp}}}] - b_2 t$. Here $t_{\mathrm{jump}}$ is the time of the jump, and $b_1$ and $b_2$ are fitting parameters respectively representing the amplitude of the jump and the pervaporation speed just after the jump.

This puzzling discrepancy between $\tau_{\mathrm{exp}}$ and $t_r$ for the fastest jumps suggests that other physical ingredients are at play in our experiments. Contact line friction is one possible such ingredient, but we have checked (see Appendix A) that it is only a correction to the model of \citet{Martinez-Calvo2020} and unable to explain our experimental dynamics.

Other sources of friction in our system originate from the air flow through the constriction and internal friction within PDMS.

 The friction of air in the narrow constriction is described using the concept of hydraulic resistance $R_h$, linking the difference in pressure to the flow rate of air with $\Delta P=R_h Q_f$. Since the constriction has a rectangular tall cross-section, we have  $R_h\simeq {12\eta_\mathrm{air} L_\mathrm{p}}/{[h w_\mathrm{p}^3(1-0.630 w_\mathrm{p}/h)]}$ when $w_{\mathrm{p}}<h$, using the viscosity of air $\eta_\mathrm{air} = 1.6 \times 10^{-5}$~m$^2 \cdot \mathrm{s}^{-1}$ \citep{Bruus2008}. From the hydraulic capacitance $C_h$ defined earlier in section \ref{Sec:compliance}, we predict the relaxation time of the flexible cavity, when friction is dominated by air, to be $\tau=R_h C_h$. A numerical application provides a value of 2 ms, also way too short.
% \item Friction of water in the output channel. Friction occurs in the main output channel $R_{h,water}=\frac{12\eta_\mathrm{water} L_\mathrm{out}}{w_\mathrm{out} h^3(1-0.630 h/w_\mathrm{out})}$ (for $h<w_\mathrm{out}$) et calculer $\tau=R_{h,water}C_h$ Ou alors  Discussion of Martinez-Calvo above. 

Concerning internal friction inside the PDMS wall, we can elaborate using results from \cite{Placet2015} showing that the complex elastic modulus for harmonic vibrations,  $E^*=E'+iE''$, presents a loss modulus $E''$ that is much smaller than the elastic modulus $E'$,  with a ratio $E''/E'<0.2$ for  frequencies smaller than 100 Hz, as is the case here. According to \citet{Landau1986}, a clamped beam of length $w_\mathrm{out}$ and thickness $\delta$ should oscillate at a complex frequency $\omega^*=\omega'+i\omega''=\frac{22.4}{w_\mathrm{out}^2}\sqrt{\frac{E^*\delta^2}{12\rho_\mathrm{PDMS}}}$, where $\rho_\mathrm{PDMS}$ is the density of the solid.
A small loss modulus thus means a natural frequency $\omega'$ much smaller than attenuation $\omega''$: the relaxation of the beam should produce noticeable underdamped oscillations. Here we clearly observe an  overdamped relaxation. As a conclusion, internal solid friction does not seem high enough to account for the relaxation time.

%[LUDOVIC ---- WE CAN ALSO PROPOSE AS POSSIBLE EXPLANATION FOR THE INCONSISTENCY THE 3D GEOMETRY OF THE MENISCUS, and the fact that when the interface gets out of the constriction, its in plane curvature decreases but its out of plane curvature gradually increases due to the deflexion of the channel. This could increase the time of release of the constriction out of the constriction. When the jump is very long, this supplemental delay is relatively negligible but when the time is short, this delay can be relatively significant. Recall that Martinez-Calvo assume a Heaviside pressure condition. In practice, an Heaviside pressure condition at the meniscus is not achievable due to the gradual change of curvature imposed by the geometry.]

\section{Summary and outlook}
\label{part:Summary}

In this study, we have proposed a novel design for biomimetic plant leaves containing constrictions and have shown that they were replicating the role of biological pits in that they enable to recover a strongly intermittent propagation of air embolism. A combination of macroscopic experiments and confocal imaging enabled us to prove that the intermittency is due to a coupling between the compliance of the channel and the capillary pressure at the interface, itself strongly dependent on the constriction geometry.

%It is interesting to discuss more in-depth the connections between our study and the behavior in real plants. 
For future studies, we seek to use this combination of a compliant microchannel and a constriction as a building block for a more complex architecture of biomimetic leaves enabling to reproduce more closely the universal dynamics reported by \citet{Brodribb2016PNAS,Brodribb2016NP} across various species of plants.

A step further towards more realist biomimetic chips would also consist in replacing PDMS by more hygrophilic materials such as hydrogels or cellulose-based porous materials. Recent studies showed for both wood \citep{Penvern2020} and cellulose-based fabrics \citep{Ma2022} that the higher water solubility of these materials leads to dynamics characterized by a most significant role played by bound water.

We anticipate that our results will have significance for a broader audience, in particular in pumpless microfluidic chips \citep{Choi2017} as well as in chemical engineering where pervaporation is largely employed \citep*{Bacchin2022}. The coupling between the compliance of the channel and the strongly variable pressure induced by the constrictions offer a unique way to passively generate controlled non-linear dynamics.

\section*{Acknowledgements}

Danièle Centanni is gratefully acknowledged for her implication in microfabrication, Delphine Débarre, Irène Wang and Sylvie Costrel for their assistance in confocal imaging, Florent Marquet for his help in the mechanical modelling, and Yves Méheust for fruitful discussions. Funding by ANR, under the grant PHYSAP ANR619-CE30-0010-02, is gratefully acknowledged, as well as our collaborators Xavier Noblin, Céline Cohen from Université Côte d'Azur, Eric Badel, José Torres-Ruiz and Hervé Cochard from INRAE Clermont-Ferrand for insightful discussions.

\begin{appendix}

\section{Effect of contact line friction on the relaxation following the jump}

In this Appendix, we refine the analysis of \citet{Martinez-Calvo2020} by accounting for the presence of contact line friction at the meniscus. Indeed, such an ingredient is present as soon as contact lines are moving \citep{Snoeijer2013}, and it significantly affects dynamical processes such as wetting or dewetting dynamics, drop sliding \citep{Winkels2011} or the damping of sloshing \citep*{Dollet2020}, to cite but a few examples.

Within the assumptions discussed in section \ref{Sec:physical_interpretation_relaxation}, \citet{Martinez-Calvo2020} showed that the pressure field in the start-up flow in a shallow deformable channel obeys a nonlinear diffusion equation. The nonlinearity originates from the cross-section-dependent hydrodynamic resistance. For the sake of simplicity, and since our experiments exhibit only moderate cross-section variations (see the snapshots of figure \ref{Fig:S_t_ref}b), we neglect this nonlinear effect and the associated channel deformation in this Appendix. The pressure field then obeys the equation:
\begin{equation} \label{Eq:diffusion_equation}
    \frac{\partial p}{\partial t} = D \frac{\partial^2 p}{\partial x^2} ,
\end{equation}
with an effective diffusion coefficient given by:
$$ D = \frac{60h^3 B}{\eta w_{\mathrm{out}}^4} , $$
and an initial condition:
\begin{equation} \label{Eq:initial_condition}
    p = -p_c ,
\end{equation}
where $p_c$ is the capillary threshold, i.e. the maximal Laplace pressure reached at the air/liquid interface. The boundary condition at the closed end of the channel $x = L_{\mathrm{out}}$ is:
\begin{equation} \label{Eq:channel_end_condition}
    \frac{\partial p}{\partial x} = 0 .
\end{equation}

\citet{Martinez-Calvo2020} considered a sudden release of the pressure from time $t = 0$, and at $x = 0$. This does not apply in our experiments, for the following reasons. First, the meniscus constitutes a moving boundary, hence the boundary condition should be applied at the time-varying location of the meniscus; we neglect this effect, since the jump length remains generally one order of magnitude shorter than the exit channel length. Second, the pressure release is not instantaneous, and it takes some time for the meniscus to equilibrate once the capillary threshold is passed. Contact line friction is one possible reason for such a delay, which we now consider.

During the post-jump relaxation, the contact angle at the meniscus is a receding contact angle $\theta_r$, which obeys Cox--Voinov law (see e.g. \citet{Snoeijer2013}):
$$ \theta_r^3 = \theta_{0r}^3 - \frac{9A\eta |\dot{L}|}{\gamma} , $$
where $\theta_{0r}$ is the equilibrium contact angle and $A$ the logarithm of the ratio of a macroscopic cut-off length and a microscopic one, which can be taken with reasonable precision as $A \approx 10$ (the precise value of $A$ will not matter in the subsequent discussion). This velocity-dependent contact angle induces a modulation of Laplace pressure. In principle, we should account for two contact angles at the glass and PDMS interfaces. We neglect this difficulty, as well as the influence of the side walls $y = \pm w_{\mathrm{out}}/2$, and we use the boundary condition:
$$ p = -\frac{\gamma}{h} \cos\theta_r = -p_{\mathrm{rem}} - \frac{\gamma}{h} (\cos\theta_r - \cos\theta_{r0}) , $$
where the remanent Laplace pressure $p_{\mathrm{rem}}$ discussed at the end of section \ref{part:confocal} has been introduced, and where $\theta_r$ may be considered as an effective contact angle. In our range of velocities, $|\theta_r - \theta_{r0}| \ll 1$, hence we can linearise the boundary condition to obtain:
$$ p \simeq -p_{\mathrm{rem}} - \frac{3A\eta\sin\theta_{r0}}{h\theta_{r0}^3} |\dot{L}| . $$
Finally, using Poiseuille law $|\dot{L}| = -(h^2/12\eta)\partial p/\partial x$, we can write the boundary condition as a mixed boundary condition:
\begin{equation} \label{Eq:mixed_condition}
    p + p_{\mathrm{rem}} = \frac{Ah\sin\theta_{r0}}{4\theta_{r0}^3} \frac{\partial p}{\partial x} .
\end{equation}

The solution of the diffusion equation (\ref{Eq:diffusion_equation}) with initial condition (\ref{Eq:initial_condition}) and boundary conditions (\ref{Eq:channel_end_condition}) and (\ref{Eq:mixed_condition}) can be found e.g. in section 4.3.6 of \citet{Crank1975}. It writes:
\begin{equation} \label{Eq:pressure_sum}
    \frac{p + p_c}{p_c - p_{\mathrm{rem}}} = 1 - \sum_{n=1}^\infty \frac{2\lambda\cos[\beta_n (1 - x/L_{\mathrm{out}})] \exp(-\beta_n^2 Dt/L_{\mathrm{out}}^2)}{(\beta_n^2 + \lambda^2 + \lambda)\cos\beta_n} ,
\end{equation}
where the coefficients $\beta_n$ are the positive roots of $\beta\tan\beta = \lambda$, and where $\lambda = 4\theta_{r0}^3 L_{\mathrm{out}}/(Ah\sin\theta_{r0})$. In the current context, this dimensionless parameter compares bulk viscous dissipation to contact line friction. To estimate its numerical value, we take as an effective contact angle $\theta_{r0} = 60^\circ$, a rough estimate from the snapshots of the meniscus; we then get $\lambda = 45$, which suggests that contact line friction remains secondary in the relaxation process. To confirm this fact, we may compute the relaxation time $\tau_r$ given by the slowest decaying exponential in (\ref{Eq:pressure_sum}): $\tau_r = L_{\mathrm{out}}^2/\beta_1^2 D = t_r/\beta_1^2$, where $t_r$ is given by (\ref{Eq:relaxation_time}). Numerical resolution yields $\beta_1 = 1.54$, whence the estimate $\tau_r = 3$~ms, way shorter than the experimental relaxation time.

\end{appendix}

\bibliographystyle{jfm}
\bibliography{Constriction_unique_2}

\end{document}